\documentclass[obeyspaces,
  journal=pasa,
  manuscript=research-paper,
  year=2025,
  volume=YY,
]{cup-journal}

\usepackage{graphicx} 
\usepackage{amsmath}
\usepackage{amssymb,microtype,siunitx,booktabs}
\usepackage[skip=0.5ex]{subcaption}
\usepackage{url}
\usepackage{hyperref} 

\newcommand{\degr}{$^\circ$}
\newcommand{\arcmin}{$'$}
\newcommand{\arcsec}{$''$}
\newcommand{\Mstellar}{$M_{\star}$}

\newcommand{\Mdyn}{~M$_{\rm dyn}$}
\newcommand{\Msun}{~M$_{\odot}$}
\newcommand{\MMsun}{M$_{\odot}$}
\newcommand{\Ha}{H$\alpha$}
\newcommand{\HI}{H\,{\sc i}}
\newcommand{\FHI}{$F_{\rm HI}$}
\newcommand{\MHI}{$M_{\rm HI}$}
\newcommand{\DHI}{$D_{\rm HI}$}
\newcommand{\RHI}{$R_{\rm HI}$}
\newcommand{\vsys}{$v_{\rm sys}$}
\newcommand{\vlg}{$v_{\rm LG}$}
\newcommand{\vrot}{$v_{\rm rot}$}
\newcommand{\vesc}{$v_{\rm esc}$}
\newcommand{\kms}{~km\,s$^{-1}$}

\title[ASKAP discovery of a bipolar outflow]{ASKAP discovery of a 30~kpc bipolar outflow from the edge-on disk of the nearby spiral galaxy ESO\,130-G012}

\author{B\"arbel S. Koribalski}
\affiliation{Australia Telescope National Facility, CSIRO, Space and Astronomy, P.O. Box 76, Epping, NSW 1710, Australia}
\alsoaffiliation{Western Sydney University, Locked Bag 1797, Penrith South DC, NSW 2751, Australia}

\email[B.S. Koribalski]{Baerbel.Koribalski@csiro.au}

\author{Roland M. Crocker}
\affiliation{Research School of Astronomy and Astrophysics, Australian National University, Canberra 2611, ACT, Australia}
\author{Ildar Khabibullin}
\affiliation{Universit\"ats-Sternwarte, Fakult\"at f\"ur Physik, Ludwig-Maximilians-Universit\"at M\"unchen, Scheinerstr.~1, D-81679 M\"unchen, Germany}
\alsoaffiliation{Max-Planck-Institut f\"ur Astrophysik, Karl-Schwarzschildstr. 1, D-85748, Garching, Germany}
\author{Anna Ivleva}
\affiliation{Universit\"ats-Sternwarte, Fakult\"at f\"ur Physik, Ludwig-Maximilians-Universit\"at M\"unchen, Scheinerstr.~1, D-81679 M\"unchen, Germany}
\author{Klaus Dolag}
\affiliation{Universit\"ats-Sternwarte, Fakult\"at f\"ur Physik, Ludwig-Maximilians-Universit\"at M\"unchen, Scheinerstr.~1, D-81679 M\"unchen, Germany}
\alsoaffiliation{Max-Planck-Institut f\"ur Astrophysik, Karl-Schwarzschildstr. 1, D-85748, Garching, Germany}
\author{Umberto Maio}
\affiliation{INAF Italian National Institute of Astrophysics -- Astronomical Observatotry of Trieste, via G.~Tiepolo~11, 34143 Trieste, Italy}
\alsoaffiliation{IFPU Institute for Fundamental Physics of the Universe, Miramare Campus, via Beirut~2, 34014 Trieste, Italy}
\author{Ralf-J\"urgen Dettmar}
\affiliation{Ruhr University Bochum, Faculty of Physics and Astronomy, Astronomical Institute (AIRUB), 44780 Bochum, Germany}
\author{Jacco Th. van Loon}
\affiliation{Lennard-Jones Laboratories, Keele University, ST5 5BG, UK}
\author{Stanislav Shabala}
\affiliation{School of Natural Sciences, University of Tasmania, Private Bag 37, Hobart 7001, Australia}

\received {17 Dec 2025}
\revised  {dd mm 2025}
\accepted {dd mm 2025}
\published{dd mm 2025}


\begin{document}

\begin{abstract}
We present the discovery of a large-scale, limb-brightened outflow, extending at least 30~kpc above and below the star-forming disk of the edge-on galaxy ESO\,130-G012 ($D$ = 16.9~Mpc). Partially obscured by Galactic foreground stars and dust, this optically unremarkable, low-mass galaxy reveals one of the largest known hourglass-shaped outflows from the full extent of its bright stellar disk. The outflow was discovered in 944~MHz radio continuum images from the Australian Square Kilometre Array Pathfinder (ASKAP) obtained as part of the "Evolutionary Map of the Universe" (EMU) project. Its height is at least 3$\times$ that of the stellar disk diameter ($\sim$10~kpc), while its shape and size most resemble the large biconical, edge-brightened FUV and X-ray outflows in the nearby starburst galaxy NGC~3079. The large-scale, hourglass-shaped outflow of ESO\,130-G012 appears to be hollow and originates from the star-forming disk, expanding into the halo with speeds close to the escape velocity before likely returning to the disk. Given ESO\,130-G012's modest star formation rate, the height of the outflow is surprising and unusual, likely made possible by the galaxy's relatively low gravitational potential. Follow-up observations are expected to detect hot gas inside the bipolar outflow cones and magnetic fields along the X-shaped outflow wings. Neutral gas may also be lifted above the inner disk by the outflow.
\end{abstract}

\section{Introduction} 
\label{sec:intro}

Galactic outflows are a key component of galaxy evolution, enriching, shocking and shaping their environments \citep[e.g.,][]{VeilleuxHawthorn2005, Veilleux2020, Tumlinson2017}. Outflow morphologies vary from bipolar lobes and bubbles to X-shaped cones and collimated jets, depending on their origin, velocity, age and driving force (e.g., active galactic nuclei, hot stellar winds, or supernova explosions). The observed outflow sizes range from less than one parsec for stellar winds to $>$1~kpc for the large-scale winds of star-forming galaxies. Here we focus on the latter. In their review, \citet{Heckman1993} define `superwinds' as {\it galactic-scale mass outflows driven by the collective effect of supernova explosions and stellar winds associated with powerful, compact starbursts.} In this context, galaxies like M\,82 \citep{Lehnert1999}, NGC~253 \citep{Pietsch2000, Strickland2002}, NGC~1808 \citep{Koribalski1993a, Koribalski1993b}, and NGC~3079 \citep{Veilleux1994, Hodges2020} are often noted as `superwind' galaxies \citep[see also][]{Strickland2004a,Strickland2004b}. 

\citet{dettmar1992} highlights the `disk-halo connection' in galaxies where star formation drives outflows not only from the nuclear region but, in some cases, from the extended stellar disk into the halo and concludes that magnetic fields play an important role \citep[see also][]{Heesen2018}. The edge-on galaxy NGC~891 is a prime example with \Ha\ filaments extending up to $\sim$4.5~kpc above the disk \citep{Dettmar1990}. \citet{Galante2024} and \citet{Irwin2024} search for the presence of radio halos in nearby, edge-on starburst galaxies, while \citet{Stein2025} find X-shaped magnetic field structure in many nearby edge-on disks. In contrast, Seyfert galaxies often show outflows from near their core. \citet{HotaSaikia2006} compare the small-scale outflow properties of 10 gas-rich Seyfert galaxies, all of which also show starburst activity and some show radio jets inside the bubbles. Maybe the most prominent example is Circinus, a nearby spiral galaxy showing bipolar radio bubbles or lobes emerging from near its Sy\,2 nucleus, perpendicular to the stellar disk to a height of $\sim$3~kpc \citep[e.g.,][]{Elmouttie1998-radio}.  

The star formation rate (SFR) in galaxies varies over time and depends on the fuel available in the disk and accreted from the interstellar medium (ISM). While generally strongest in the nuclear regions, star formation in galaxies spreads over the whole stellar disk, driving gas outflows into the galaxy halo and, in some cases, well beyond the disk into the circumgalactic medium \citep[CGM,][]{Tumlinson2017}. For observational and theoretical reviews of stellar-driven galactic winds see \citet{Rupke2018-review} and \citet{Zhang2018-review}, respectively. \\

In this paper, we focus on the nearby, edge-on galaxy ESO\,130-G012 and its newly discovered bipolar outflow. The ASKAP observations are described in Section~2, followed by our results in Section~3. Possible formation scenarios are discussed in Section~4, our summary and conclusions are given in Section~5 and an outlook in Section~6. 

\begin{figure}[htb] 
\centering
  \includegraphics[width=8.5cm]{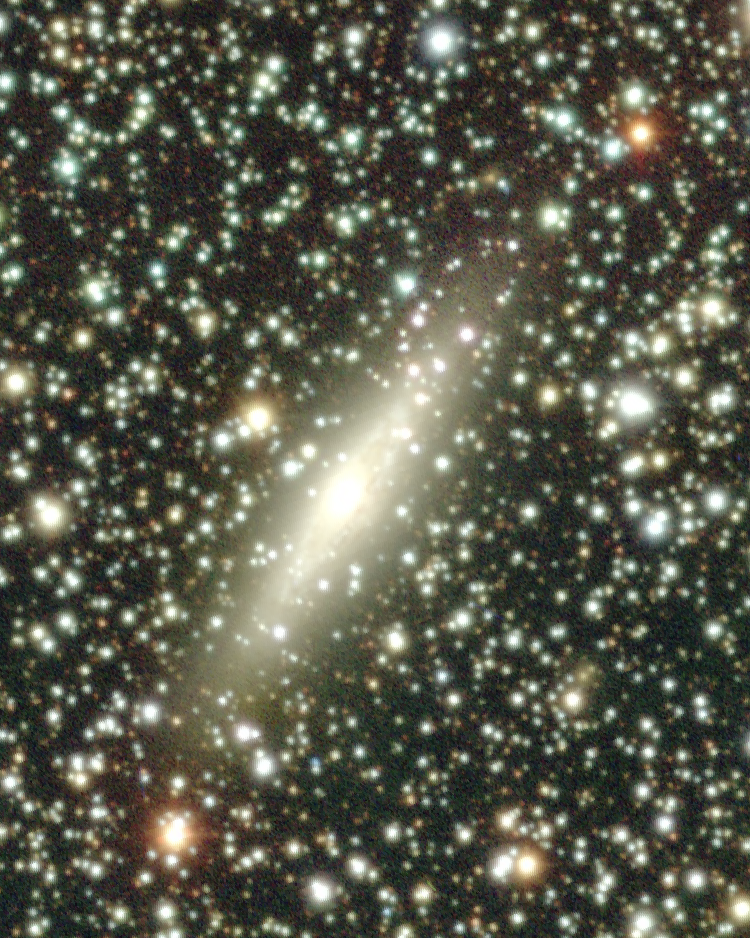} 
\caption{Optical DECaPS colour-composite image of ESO\,130-G012. The stellar disk clearly stands out despite the high density of foreground stars from the Milky Way Galactic Plane, and a curved dust lane is just visible against the bright galaxy bulge, suggesting a disk inclination of $\sim$80\degr. The ESO $B_{\rm 25.5mag}$-band diameter of the stellar disk is $\sim$1.8~arcmin corresponding to 8.8~kpc for a distance of $D$ = 16.9~Mpc. --- North is up and East to the left.}
\label{fig:decaps}
\end{figure}

\begin{table}
\centering
\caption{Properties of the galaxy ESO\,130-G012 for $D = 16.9$~Mpc}
\begin{tabular}{llc}
\hline
  source names & ESO\,130-G012 \\
  & HIZOA~J1222--58 \\
  & ZOAG G299.18+04.04 \\
  & WKK 0957 \\
  & \multicolumn{2}{l}{2MASX J12223838--58365982} \\ 
  & \multicolumn{2}{l}{WISEA J122238.37--583659.8} \\
  & IRAS~12199--5820	 \\
  & 1eRASS J122238.4--583658 \\
  & GLEAM-300 J122238--583710 \\
\hline
 $\alpha,\delta$(J2000) position & $12^{\rm h}\,22^{\rm m}\,38.4^{\rm s}$, --58\degr\,36\arcmin\,59.8$''$ & S06 \\
 Galactic $l,b$ position & 299.1805\degr, +4.0453\degr	\\
  morph. type & SB0$^+$(r)? & RC3 \\
              & S(r)a: & B95 \\
  optical $B_{\rm 25.5mag}$ size & $1.8' \times 0.4'$, $PA$ = 147\degr & ESO \\ 
  & 8.8 kpc $\times$ 2.0 kpc \\
  inclination & $\sim$80\degr\  \\ 
  2MASS $K_s$-band size & $2.1' \times 0.7'$, $PA$ = 150\degr & S06 \\ 
  & 10.2~kpc $\times$ 3.5~kpc \\
  inner ring size & $0.68' \times 0.20'$, $PA$ = 150\degr & B95 \\
  ~~~ (dust \& \Ha) & 3.3~kpc $\times$ 1.0~kpc \\
\hline
  Parkes \HI\ properties & & here \\
\hline
  \HI\ systemic velocity (\vsys) & $1533 \pm 3$\kms \\
  Local Group velocity (\vlg) & 1270\kms \\
  $->$ distance ($v_{\rm LG}/H_{\rm o}$) & 16.9~Mpc \\
  20\% \HI\ velocity width & $400 \pm 9$\kms  \\
  50\% \HI\ velocity width & $370 \pm 6$\kms  \\
  $->$ rotational velocity (\vrot) & $188 \pm 6$\kms \\
  $->$ escape velocity (\vesc) & $266 \pm 6$\kms \\
  \HI\ flux density (\FHI) & $7.9 \pm 1.2$ Jy\kms \\
  $->$ \HI\ mass (\MHI)& 5.3 ($\pm$0.8) $\times\ 10^8$\Msun \\
  $->$ \HI\ disk diameter (2\RHI) & $\sim$13~kpc at 1\Msun\,pc$^{-2}$ \\
  $->$ $M_{\rm dyn}$ for $R_{\rm HI}$ = 6.5~kpc & 5.3 ($\pm$0.5) $\times\ 10^{10}$\Msun \\
\hline
ASKAP 944~MHz properties & & here \\
\hline
  total radio flux (disk+outflow) & $65 \pm 5$ mJy (see also Table~\ref{tab:radio-prop}) \\
  total radio power & $\sim$2.2 $\times\ 10^{21}$ W\,Hz$^{-1}$ \\
\hline
infrared properties \\
\hline
 $K$-band magnitude & $8.610 \pm 0.024$ & W14 \\
 black hole mass & 5.0 ($\pm$ 2.5) $\times 10^7$\Msun & CB10 \\
 WISE W1, 2, 3, 4 g-mags & 
   8.215, 8.206, 6.344, 4.47 & C21 \\
 $->$ stellar mass (\Mstellar) & $1.1 \times 10^{10}$\Msun\ (based on W1) \\
 $->$ SFR & 0.20, 0.24 (based on W3, W4) \\
 WISE colours & W1--W3 = 1.87, W1--W2 = 0.01 \\
 WISE colour diagnostics & SF spiral, no dominant AGN \\
 IRAS 60$\mu$m, 100$\mu$m & 2.37~Jy, 5.63~Jy & Y93 \\ 
 $->$ FIR luminosity & 5.1 ($\pm$1.1) $\times 10^{42}$ erg\,s$^{-1}$ \\
 AKARI 90$\mu$m & $3.463 \pm 0.141$ Jy & Y09 \\
\hline
\hline
\end{tabular}
\flushleft {References: S06 \citep{Skrutskie2006}, RC3 \citep{deVaucouleurs1991}, B95 \citep{Buta1995}, ESO \citep{Lauberts1982}, W14 \citep{Williams2014}, CB10: \citep{CB2010}, C21 \citep{Cutri2021}, Y93 \citep{Yamada1993}, Y09 \citep{Yamamura2009}.}
\label{tab:properties}
\end{table}

\begin{figure*} 
\centering
 \includegraphics[width=16cm]{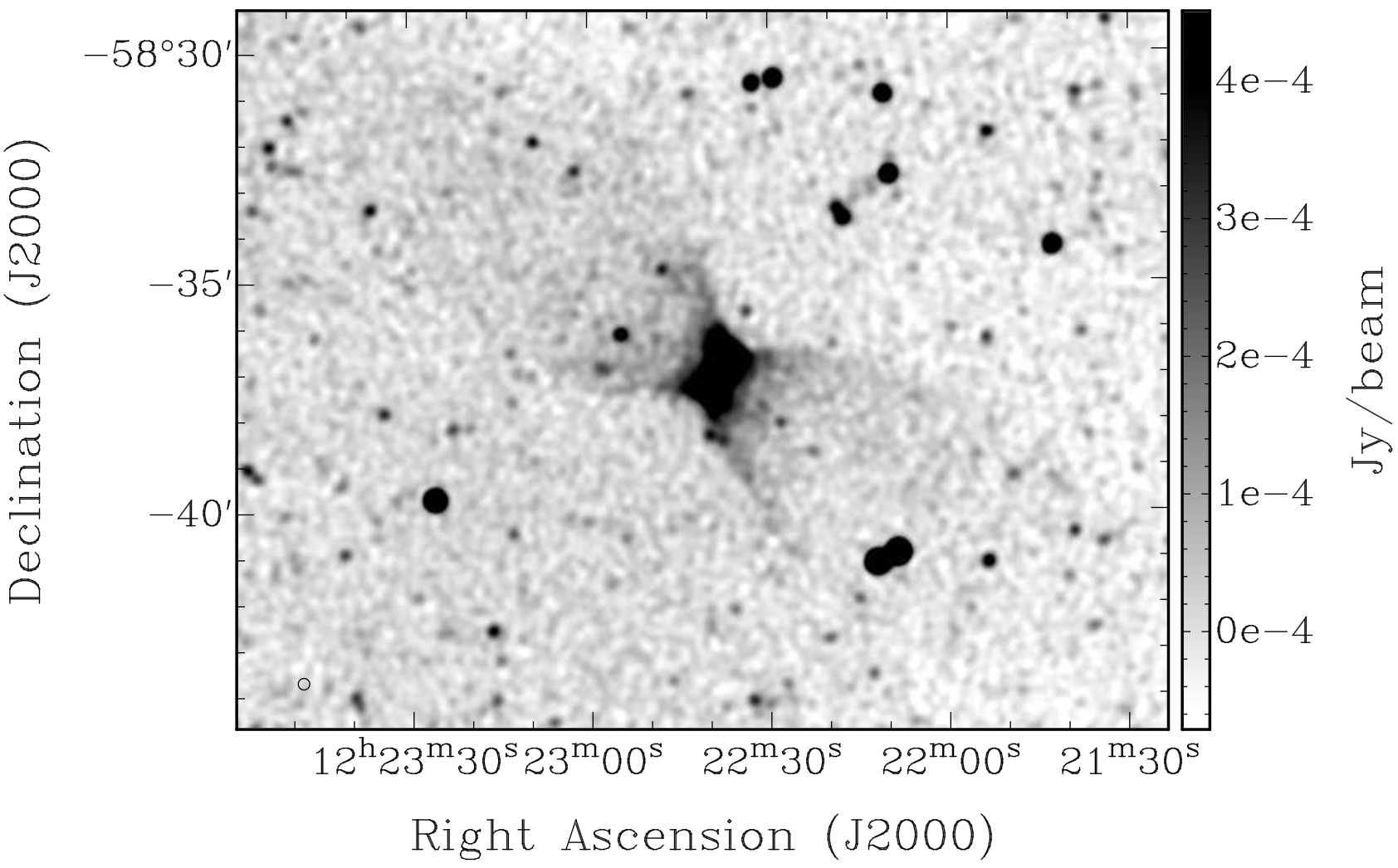}
\caption{ASKAP EMU 944~MHz radio continuum image of the spectacular outflow from the edge-on galaxy ESO\,130-G012. At the galaxy distance of 16.9~Mpc, the detected outflow height of at least 6\arcmin\ corresponds to $\sim$30~kpc (see also Figure~\ref{fig:bubble-overlays}). The ASKAP resolution of 15\arcsec\ is indicated in the bottom left corner. }
\label{fig:emu}
\end{figure*}

\section{ASKAP Observations and Data Processing}
\label{sec:obs}

ASKAP is a powerful 6.4-km diameter radio interferometer consisting of 36 $\times$ 12-m antennas, dedicated to large radio sky surveys \citep{Johnston2008, Norris2011, Koribalski2012, Koribalski2020, McConnell2020, Norris2021, Duchesne2023} plus guest projects. Each antenna is equipped with an innovative, wide-field Phased Array Feed \citep[PAF,][]{Chippendale2015} which operates at frequencies from 0.7 to 1.8 GHz and delivers a $\sim$30 deg$^2$ field of view out to the half power point. The currently available ASKAP bandwidth of 288~MHz is divided into $288 \times 1$ MHz coarse channels. For a comprehensive system overview see \citet{Hotan2021}. A brief summary of ASKAP technical and science highlights is presented in \citet{Koribalski2022}. \\
 
Our field of interest was observed on the 14th of April 2024 as part of the EMU project, with the band centred at 943.5~MHz (scheduling block 61083). The total integration time was 10~hours and the field centre is at $\alpha,\delta$(J2000) = $12^{\rm h}\,20^{\rm m}\,34^{\rm s}$, --60\degr\,19\arcmin\,18\arcsec. ASKAP PAFs were used to form 36 beams arranged in a $6 \times 6$ closepack footprint. The data processing was done in the ASKAPsoft pipeline \citep{Wieringa2020, EMU-PS}. We obtained fully calibrated ASKAP EMU radio continuum images from the CSIRO ASKAP Science Data Archive (CASDA)\footnote{CASDA: \url{https://data.csiro.au/domain/casdaObservation}}. We use the convolved ASKAP images with an angular resolution of 15\arcsec\ as well as the high-resolution ASKAP images with a synthesized beam of $7.9'' \times 7.3''$. We measure an rms noise of $\sim$22~$\mu$Jy\,beam$^{-1}$ near ESO\,130-G012 in the 15\arcsec-resolution images. 

We also use the clean component residual maps, which are available in CASDA under the `Ancillary' tab. These turn out to very useful to examine the faint diffuse emission of extended sources as bright sources are subtracted out.

\begin{table}[]
\centering
\begin{tabular}{cccc}
\hline
  outflow shape & elongated bipolar bubbles \\
  outflow height & $\gtrsim$6\arcmin\ (30~kpc) on each side \\ 
  largest outflow width & $\sim$5.7\arcmin\ (28~kpc) at 25~kpc height \\
  outflow base (waist) & $\sim$2\arcmin\ (10~kpc) \\
  outflow opening angle$^\star$ & $\sim$30\degr \\
  morphology above the base & X-shaped wings \\
  bubble appearance & limb-brightened (hollow) \\
\hline
\end{tabular}
\flushleft $^\star$The outflow opening angle is determined with respect to the vector orthogonal to the disk from the midplane on each side.
\caption{Outflow properties of ESO\,130-G012}  
\label{tab:outflow-properties}
\end{table}

\section{Results}
\label{sec:results}

\begin{figure*}[htb] 
\centering
    \includegraphics[width=14cm]{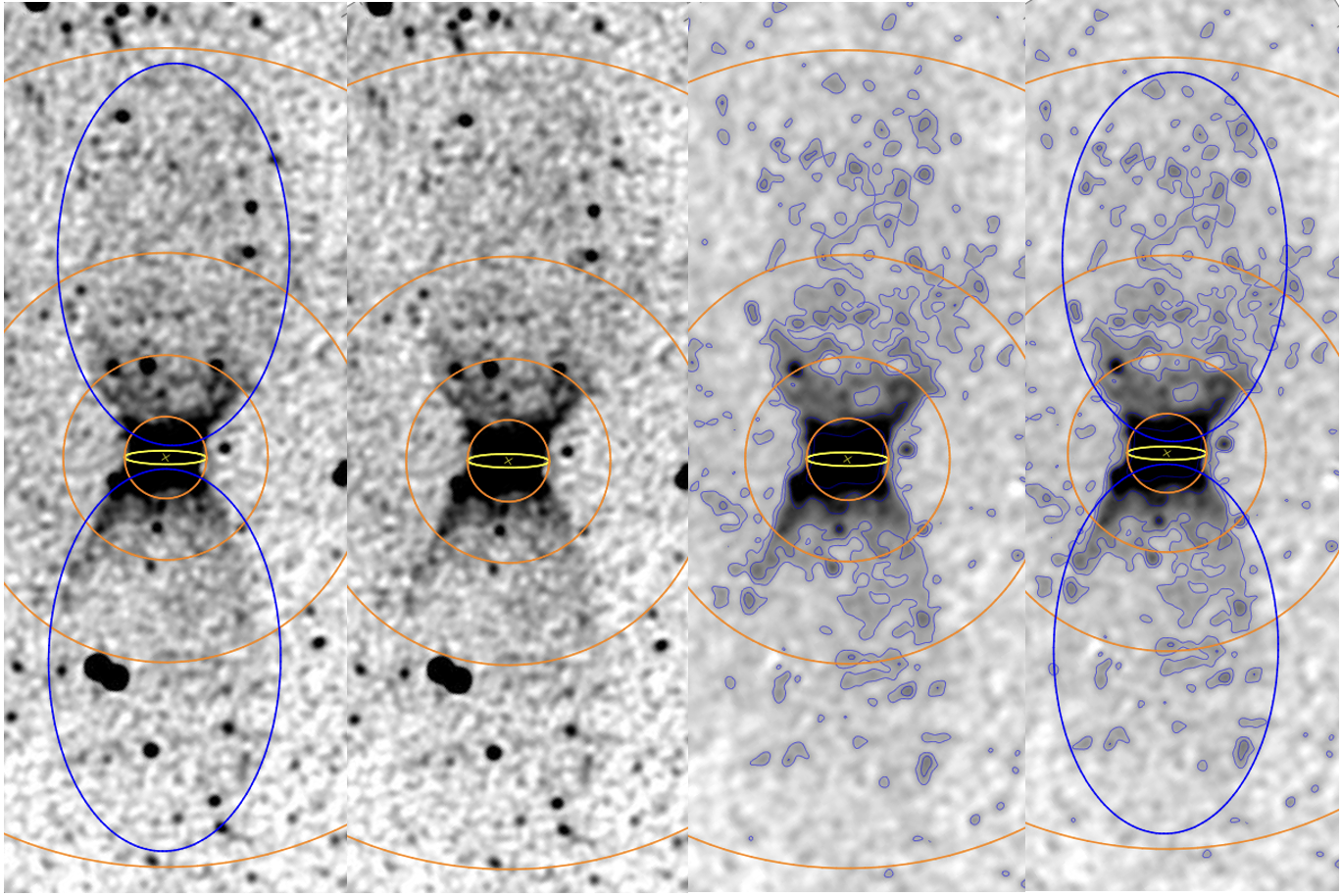}
\caption{ASKAP EMU 944~MHz radio continuum images of the bipolar outflow from the galaxy ESO\,130-G012, rotated such that the flow is approximately along the y-axis. {\bf -- Left pair:} Total intensity images at 15\arcsec\ resolution. {\bf -- Right pair:} Clean residual images smoothed to 20\arcsec\ resolution. The contour levels are 30, 60, 150 and 300~$\mu$Jy\,beam$^{-1}$. The yellow ellipse indicates the approximate size of the stellar disk. Orange circles are drawn at radii of 60\arcsec, 150\arcsec, 300\arcsec\ and 600\arcsec. At the galaxy distance of 16.9~Mpc, 60\arcsec\ corresponds to $\sim$5~kpc. The blue ellipses have sizes of 560\arcsec\ $\times$ 340\arcsec\ (46~kpc $\times$ 28~kpc).}
\label{fig:bubble-overlays}
\end{figure*}

\begin{figure*}[htb] 
\centering
  \includegraphics[width=15cm]{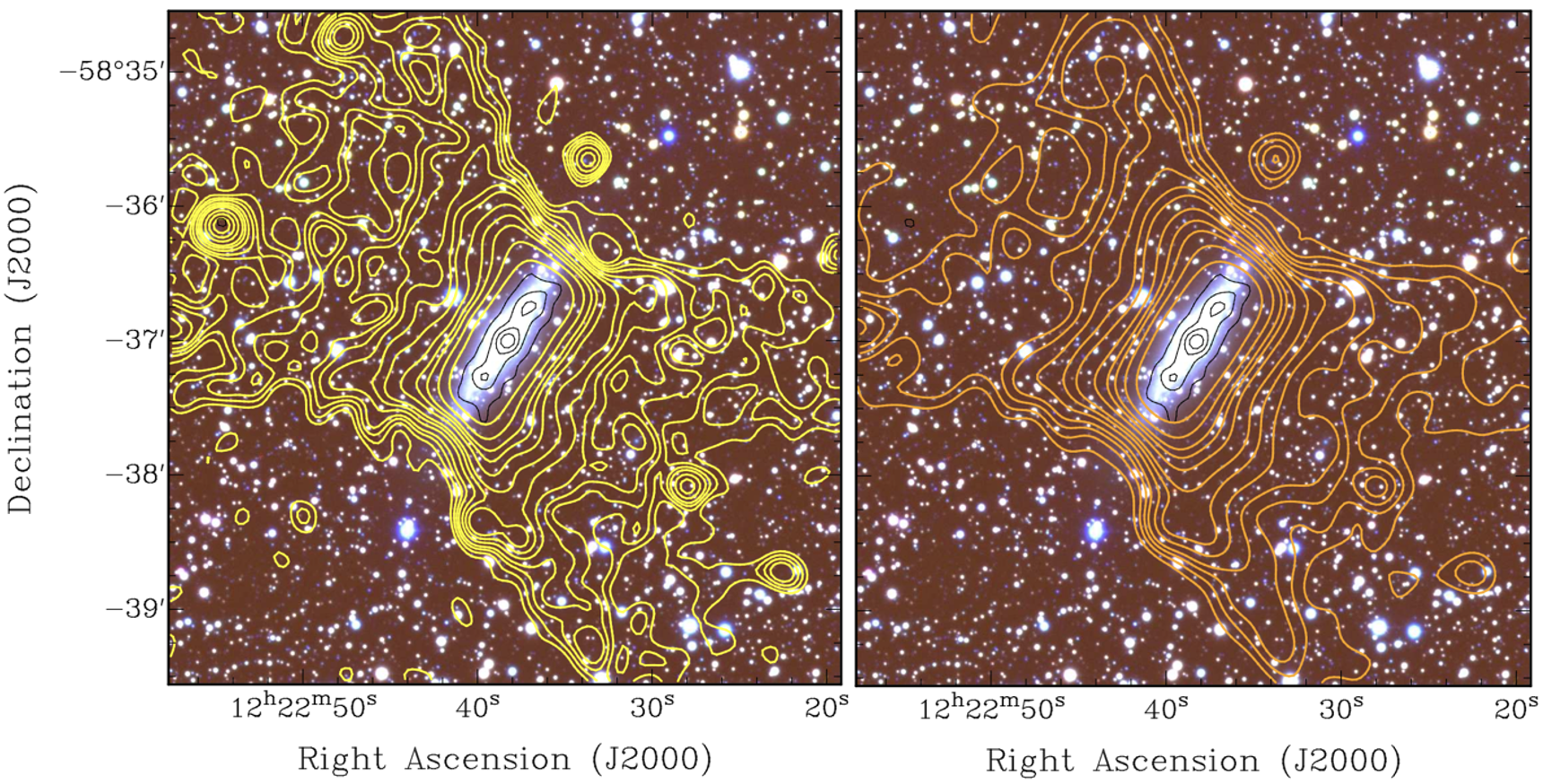} 
\caption{Zoom-in of the ESO\,130-G012 outflow, here shown to a height of $\sim$160\arcsec\ ($\sim$13~kpc). --- DECaPS $zrg$-band optical image overlaid with ASKAP EMU 944~MHz radio continuum contours. {\bf -- Left:} Total intensity contours (yellow, 15\arcsec\ resolution). {\bf -- Right:} Clean residual contours (orange, 20\arcsec\ resolution). The levels are 0.06, 0.09, 0.12, 0.15, 0.20, 0.25, 0.3, 0.4, 0.5, 0.6, 0.8 and 1.2 mJy\,beam$^{-1}$ (yellow / orange) plus 0.5, 1, 1.4, and 1.8 mJy\,beam$^{-1}$ (black, 8\arcsec\ resolution). }
\label{fig:optical+emu}
\end{figure*}

While inspecting deep ASKAP EMU 944~MHz radio continuum images, we discovered a bipolar outflow extending at least 6\arcmin\ ($\sim$30~kpc) above and below the edge-on stellar disk of ESO\,130-G012, a nearby spiral galaxy at a distance of only 16.9~Mpc (see Table~\ref{tab:properties}). Figure~\ref{fig:decaps} shows a deep optical image of the galaxy from the Dark Energy Camera Galactic Plane Survey \citep[DECaPS,][]{Saydjari2023}\footnote{http://decaps.skymaps.info/}. Despite the high density of Galactic foreground stars, ESO\,130-G012's stellar disk (size $\sim$ 2.5\arcmin) and the curved inner dust lane around the central bulge are discernible. We estimate a disk inclination of $\sim$80\degr. The cataloged 2MASS $K_s$-band major axis diameter \citep{Jarrett2000, Skrutskie2006} is 2.1\arcmin\ at the 20~mag\,arcsec$^{-2}$ isophote. The radio emission of ESO\,130-G012, including its edge-on disk and the spectacular, limb-brightened outflow, are shown in Figures~\ref{fig:emu} -- \ref{fig:halpha}. The shape of the bipolar outflow, which may extend up to $\sim$50~kpc on each side, resembles a closed hourglass with a $\sim$10~kpc wide waist. A summary of the galaxy properties is given in Table~\ref{tab:properties}.

\subsection{Optical properties}
ESO\,130-G012 is a nearly edge-on, nearby spiral galaxy located at a Galactic latitude of only 4\degr, making optical imaging difficult. An optical colour-composite image from DECaPS, shown in Figure~\ref{fig:decaps}, reveals the galaxy's bright stellar disk, inner ring and central bulge, all partially hidden behind a multitude of Galactic foreground stars.  \citet{deVaucouleurs1991} give a morphological type of SB0$^+$(r)?, while \citet{Buta1995} give S(r)a: and note an inner ring (r) with dimensions $0.68' \times 0.20'$ (3.3 kpc $\times$ 1.0 kpc). The size of the ESO\,130-G012 stellar disk, as revealed by the DECaPS images, is $\sim$2.5\arcmin\ corresponding to $\sim$12.3~kpc, slightly larger than the optical and infrared diameters listed in Table~\ref{tab:properties}. 

ESO\,130-G012 was also observed as part of the VST Photometric \Ha\ Survey of the Southern Galactic Plane and Bulge \citep[VPHAS+,][]{Drew2014}, which provides $u, g, r, i$ and \Ha\ images at $\sim$1\arcsec\ angular resolution. The above mentioned dusty inner ring is detected in \Ha, as shown in Figure~\ref{fig:halpha}, suggesting it is star-forming. Such inner rings are often observed in spiral galaxies and are in many cases associated with bars.

\subsection{Radio properties}

\begin{figure*}[htb] 
\centering
  \includegraphics[width=11cm]{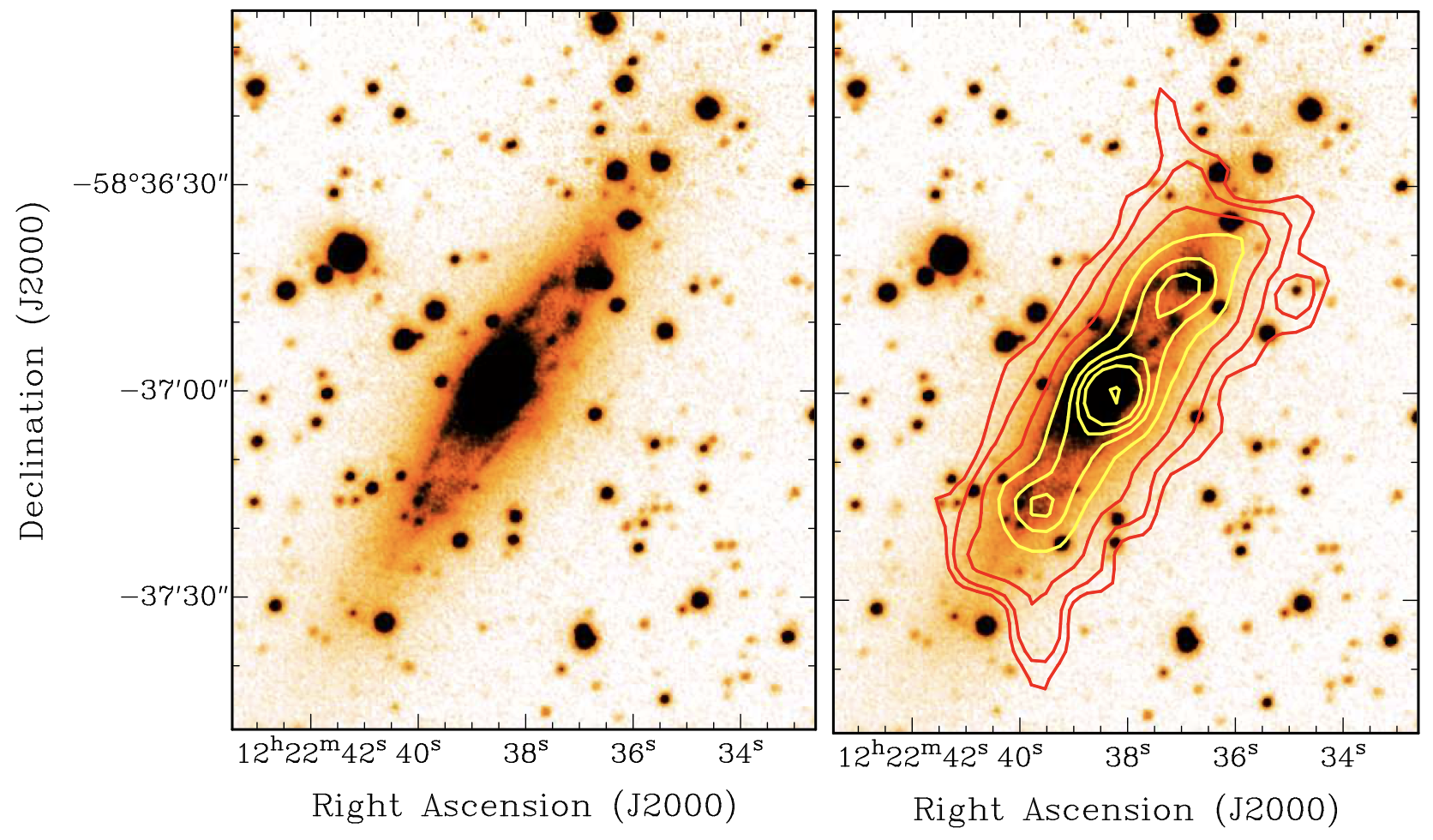} 
  \includegraphics[width=5.8cm]{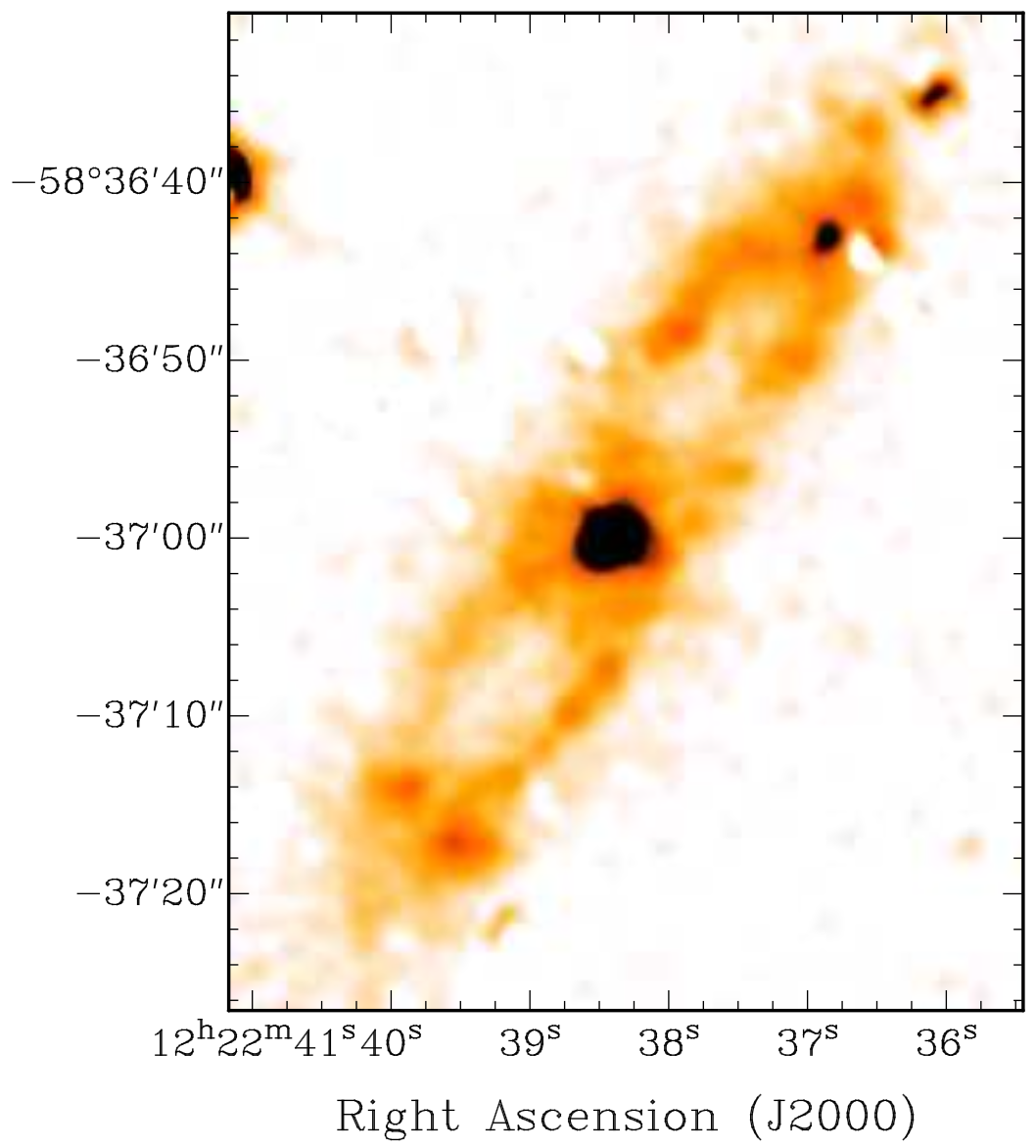} 
\caption{The inner ring and disk of ESO\,130-G012. VPHAS+ \Ha\ image (left), overlaid with ASKAP EMU 944~MHz radio continuum contours at $\sim$8\arcsec\ resolution (middle), and a zoom-in VPHAS+ continuum subtracted (\Ha-$r$) image (right). Contours in the middle image are at 0.3, 0.4, 0.6 (red), 1.0, 1.2, 1.4, 1.6 and 2.1 mJy\,beam$^{-1}$ (yellow).}
\label{fig:halpha}
\end{figure*}

The ASKAP EMU 944~MHz radio continuum emission associated with the galaxy ESO\,130-G012 consists of several components: a core, inner knots (associated with the inner ring, see Figure~\ref{fig:halpha}), a thin disk, a box-shaped thick disk, and X-shaped radio wings emerging from the edges of the box-shaped emission (see Figures~\ref{fig:bubble-overlays} \& \ref{fig:optical+emu}), forming the wide base of a huge hourglass-shaped radio outflow  which reaches at least 30~kpc above and below the disk. The outflow properties are summarized in Table~\ref{tab:outflow-properties}, and the radio continuum flux estimates for different regions are listed in Table~\ref{tab:radio-prop}. Estimates of the star formation rate (SFR) from both radio and infrared data are given in Section~3.3.1. \\

Figures~\ref{fig:emu} \& \ref{fig:bubble-overlays} highlight the size, shape and limb-brightened nature of the bipolar radio continuum outflow. Very faint radio emission can detected to nearly 50~kpc, most clearly seen in the clean residual images (Figure~\ref{fig:bubble-overlays}, right pair), with the north-eastern outflow slightly brighter than its south-western counterpart. We measure a total radio continuum flux density of $\sim$65~mJy, with half of the flux coming from the box-shaped area around and aligned with the galaxy disk ($\sim$3\arcmin\ $\times$ 2\arcmin, i.e. 15~kpc $\times$ 10~kpc, see Figure~\ref{fig:optical+emu}).

\begin{table}[htb] 
\centering
\begin{tabular}{ccccc}
\hline
  region     & flux density & comments \\
\hline
  core       &  2.1 mJy & 8\arcsec\ image \\
  core + inner ring & 16 mJy & 8\arcsec\ image \\
  inner disk & 23 mJy & 8\arcsec\ image \\ 
  disk + lower outflow & 50 mJy & $<$2\arcmin\ disk height \\ 
  upper outflow & $\sim$15 mJy & $>$2\arcmin\ disk height \\ 
\hline
  total      & $65 \pm 5$ mJy \\
\hline
\hline
\end{tabular}
\caption{Radio continuum flux estimates for different regions in the galaxy ESO\,130-G012 from the ASKAP 944~MHz images.}
\label{tab:radio-prop}
\end{table}

\vspace{0.5cm}

In the high-resolution (8\arcsec) radio continuum image we measure a core flux of 2.1~mJy. The flux densities within the lowest red and yellow contours in Figure~\ref{fig:halpha} (middle) are 23 and 13~mJy, respectively. The lowest red contour shows the first hints of wings / ears at both sides. For the core and inner ring together we measure 16~mJy. Integrating the flux up to a height of $\sim$2\arcmin\ above the disk (see Figure~\ref{fig:optical+emu}), which includes the bright, box-shaped central area and X-shaped wings, gives 50~mJy. Beyond a height of $\sim$2\arcmin\ the detected flux in the outflow is only $15 \pm 5$~mJy. See Table~\ref{tab:radio-prop} for a summary of the flux densities. The outflow possibly extends to 10\arcmin\ (50~kpc) above and below the disk of ESO\,130-G012, but deeper high-resolution radio images (with good sensitivity to low surface brightness emission) are needed to confirm this. \\

The bi-polar outflow of ESO\,130-G012 is symmetric and edge-brightened, with a prominent (closed) hourglass shape. In Figure~\ref{fig:bubble-overlays} we show ASKAP total intensity and clean residual images. The blue ellipses each have sizes of 46~kpc (height) $\times$ 28~kpc (width). The stellar disk, indicated by a yellow ellipse of $\sim$10~kpc, forms the base of the outflow. The latter is also evident in Figure~\ref{fig:optical+emu} as well as box-shaped emission in the lower outflow above and below the disk. Prominent X-shapes wings, which are seen in several other nearby, edge-on galaxies with outflows, indicate that most of the emission resides on the surface of a largely hollow structure. 

\subsubsection{Polarization and spectral index estimates}
Analysis of the polarization data from the same ASKAP observation shows no detection of polarized emission in the galaxy disk, suggesting a polarization fraction lower than 1\%. Depolarization is likely high both by Galactic foreground emission and internally; see, e.g., \citet{Stein2025}. The EMU in-band spectral index within the inner disk (where the radio flux is $>$3~mJy) is $\sim$0.57 $\pm$ 0.11, typical for star-forming galaxies \citep{Marvil2015, Klein2018}. Furthermore, we note that ESO\,130-G012 is cataloged in GLEAM-300 \citep{Duchesne2025} with an integrated flux density of $71 \pm 13$~mJy and a source size of 184\arcsec\ ($\pm$13\arcsec) $\times$ 129\arcsec\ ($\pm$8\arcsec). The EMU 944~MHz flux within the above area is $\sim$50~mJy (disk + lower outflow, see Table~\ref{tab:radio-prop}), suggesting a flat spectral index of $\alpha \approx -0.30 \pm 0.15$. The local noise (10.3~mJy) and low resolution (150.3\arcsec $\times$ 135.4\arcsec) of GLEAM-300 make this a very preliminary estimate for the disk and lower outflow combined. For example, the detected 888~MHz flux density of ESO\,130-G012 in the first data release of the Rapid ASKAP Continuum Survey \citep[RACS,][]{McConnell2020, Hale2021} is only  $\sim$30~mJy due to the survey's shallow nature (rms $\sim$0.3 mJy\,beam$^{-1}$, 25\arcsec\ resolution), which leads to $\alpha \approx -0.8 \pm 0.2$. \\

Deep, wide-band data are needed to map the spectral index of the disk and detect polarized emission in the X-shaped wings and bipolar outflow. The upgraded Australia Telescope Compact Array (ATCA) with its brand-new BIGCAT correlator, which recently replaced the Compact Array Broad-band Backend \citep[CABB,][]{Wilson2011}, would be ideal for this purpose, particularly the sensitive and nearly RFI-free 4-cm band, as well as MeerKAT UHF- and L-band follow-up observations.

\begin{figure}[htb] 
\centering
    \includegraphics[width=7cm]{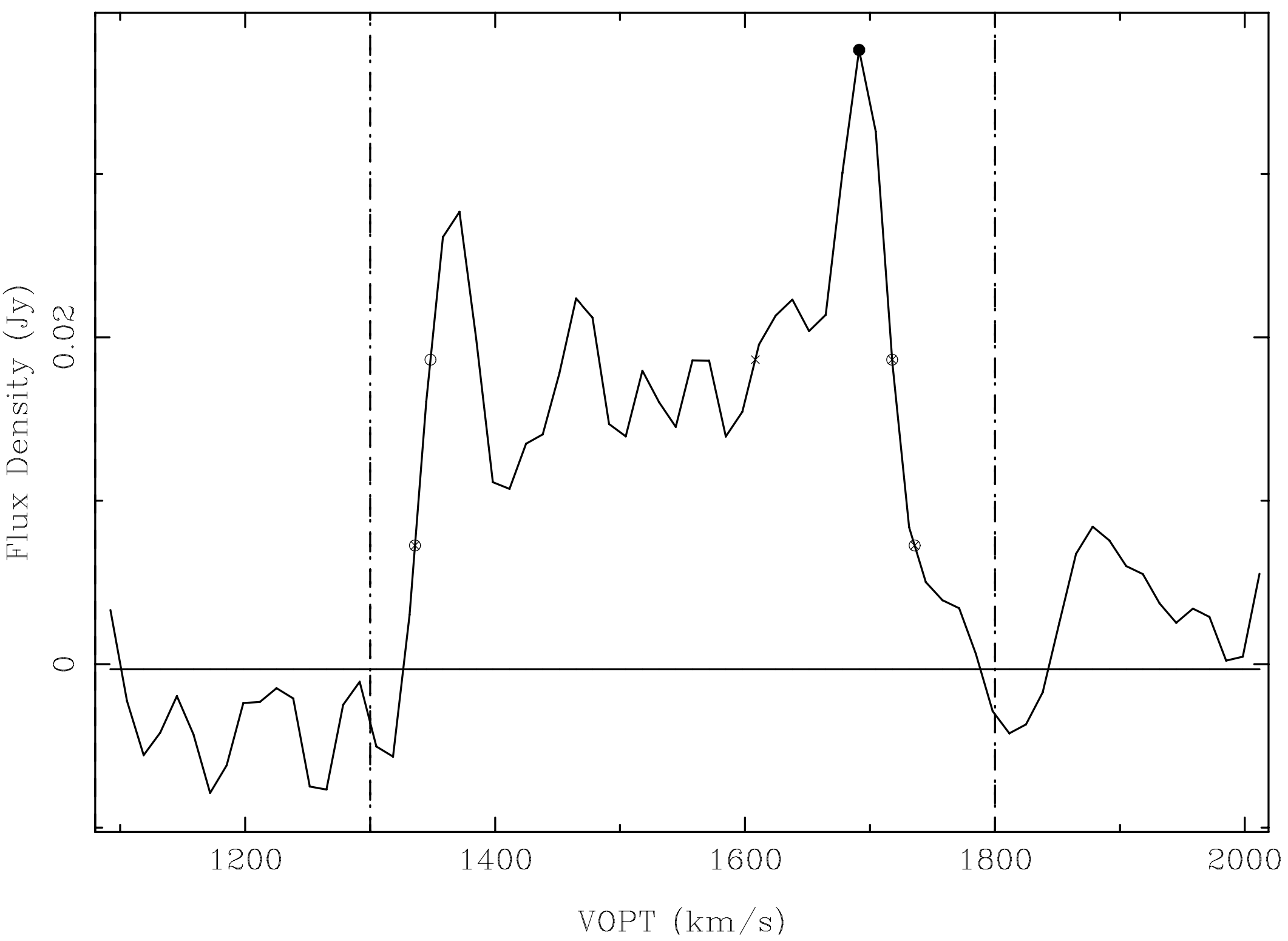}
    \includegraphics[width=8cm]{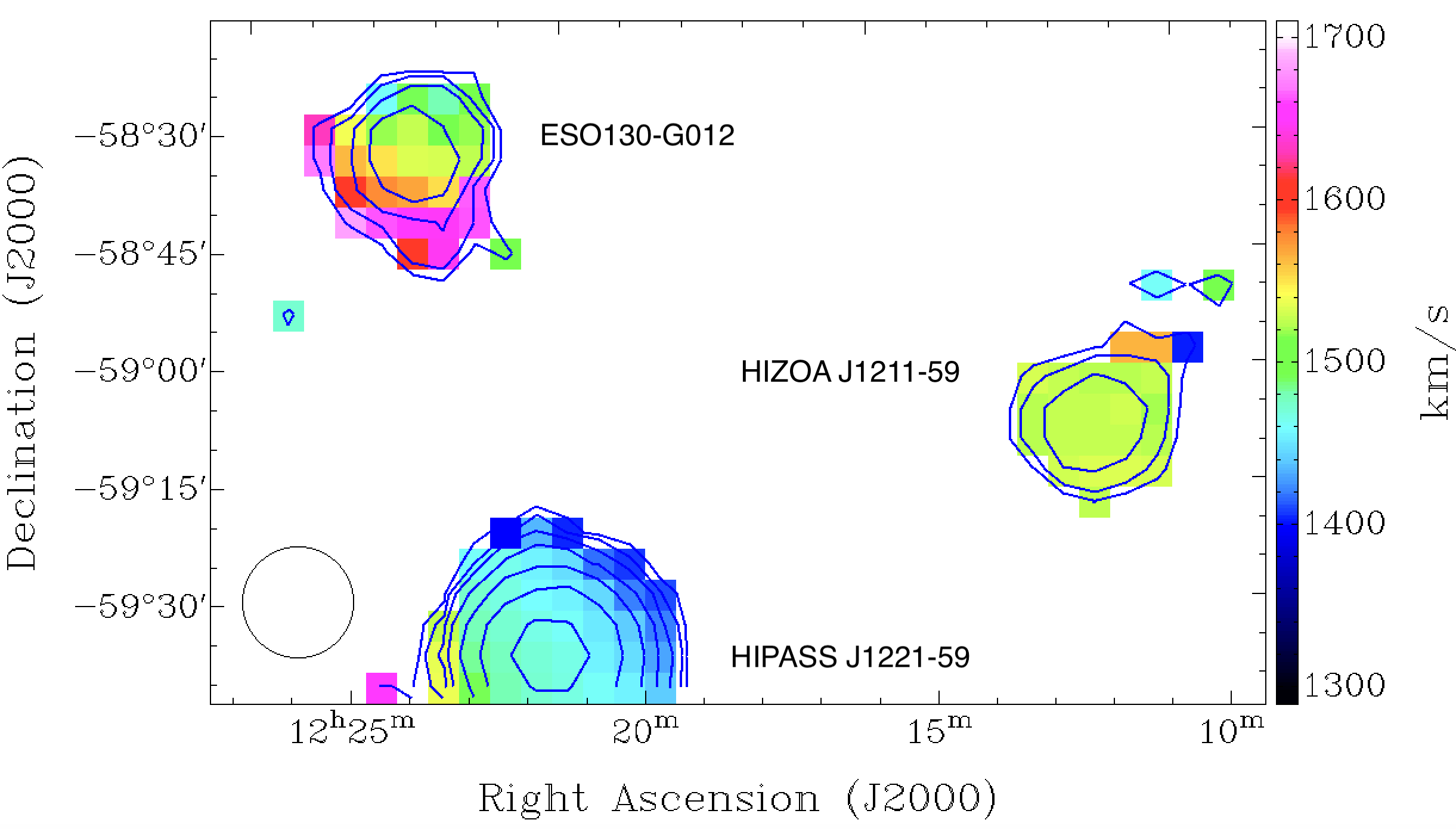}
\caption{{\bf -- (Top:)} Parkes \HI\ spectrum of the galaxy ESO\,130-G012. {\bf -- (Bottom:) } Mean \HI\ velocity field of ESO\,130-G012 and two neighboring galaxies. The gridded Parkes beam (15.5\arcmin) is shown in the bottom left corner.}
\label{fig:HI}
\end{figure}

\subsection{Infrared properties}

From the infrared magnitudes listed in Table~\ref{tab:properties} we estimate the stellar mass, \Mstellar, of ESO\,130-G012, correcting for dust extinction \citep[E($B-V$) = 0.5,][]{SF2011}. \citet{Williams2014}, using deep near infrared (NIR) photometry of \HI-detected galaxies in the Zone of Avoidance, give an extrapolated total $K$-band magnitude of $8.610 \pm 0.024$ for ESO\,130-G012, corresponding to \Mstellar\ = 2.1 ($\pm$0.1) $\times 10^{10}$\Msun. Converting from a Salpeter to a \citep{Chabrier2003} initial mass function (IMF) results in \Mstellar\ = 1.2 ($\pm$0.1) $\times 10^{10}$\Msun.
Alternately, following \citet{Jarrett2023-Mstellar} and \citet{Cluver2025}, who adopt a Chabrier IMF, we estimate \Mstellar\ = $1.1 \times 10^{10}$\Msun\ based on the extinction-corrected WISE W1 magnitude. 

\subsubsection{Star formation rate (SFR)}

We estimate the disk SFR of ESO\,130-G012 from (a) the ASKAP 944~MHz radio continuum emission, (b) the IRAS FIR emission, and (c) the WISE W3- and W4-band magnitudes. 

(a) The formation rate of recent, high-mass stars ($M > 5$\Msun) is calculated using SFR (\MMsun\,yr$^{-1}$) = 0.03 $D^2$ $S_{\rm 20cm}$ \citep{Condon2002}, where $D$ is the distance in Mpc and $S_{\rm 20cm}$ the 20-cm radio continuum flux density in Jy. We measure $S_{\rm 944MHz} \sim 23$~mJy for the inner disk (see Table~\ref{tab:radio-prop}), resulting in SFR $\sim$0.2\Msun\,yr$^{-1}$. In order to derive the formation rate of all stars ($M > 0.1$\Msun) we multiply this value by 5.5 \citep[see][]{Condon2002}, resulting in $\sim$1.1\Msun\,yr$^{-1}$. Adjusting from a Salpeter to a Chabrier IMF results in SFR $\sim$ 0.6\Msun\,yr$^{-1}$ for the disk of ESO\,130-G012. 
                                
(b) Using the IRAS flux densities \citep[see][]{Yamada1993}, we derive $L_{\rm FIR}$ = 5.1 ($\pm$ 1.1) $\times 10^{42}$~erg\,s$^{-1}$ 
and SFR$_{\rm FIR} = 0.23$\Msun\,yr$^{-1}$ based on 
\citet{Kennicutt1998}
and a Salpeter IMF. Adjusting to a Chabrier IMF gives
0.13\Msun\,yr$^{-1}$. 

(c) Following \citet{Cluver2025}, who use the Chabrier IMF, we estimate SFR = 0.20\Msun\,yr$^{-1}$ and  0.24\Msun\,yr$^{-1}$ based on the extinction-corrected WISE W3- and W4-band magnitudes, respectively. 

In summary, we find SFR values of 0.1 -- 0.6\Msun\,yr$^{-1}$ for the disk of ESO\,130-G012. We estimate an \HI\ depletion time of \MHI/SFR $\sim$ 1 -- 5~Gyr.

\subsubsection{Radio-FIR correlation}
We find that the disk of ESO\,130-G012 lies on the radio-FIR correlation, derived for star-forming galaxies by \citet{Helou1985}. For the inner disk, where we measure a flux density of 23~mJy (see Table~\ref{tab:radio-prop}), the logarithmic FIR/radio flux density ratio, $q_{\rm FIR}$, is 2.24, close to the typical value of 2.34 \citep{Yun2001}. For the total flux of 65~mJy, which includes the bipolar outflow, we get a much lower value of $q_{\rm FIR}$ = 1.78.

\subsection{X-ray properties}

The SRG/eROSITA All-sky Survey catalog (eRASS1) reveals a point-like source in the centre of ESO\,130-G012 with an absorbed 0.2 -- 2.3~keV flux of 5.8 ($\pm$1.6) $\times 10^{-14}$ erg\,s$^{-1}$\,cm$^{-2}$ \citep{Merloni2024}, so there might be a low-luminosity ($\sim$3.4 $\times 10^{39}$~erg\,s$^{-1}$) active galactic nucleus (AGN)
or several X-ray binaries in the galaxy center. Based on the total \HI\ column density in the galaxy direction of $\sim$1.3 $\times 10^{21}$~cm$^{-2}$, we used a correction factor of unabsorbed/absorbed = 1.7 for the 0.2 -- 2.3~keV band and a powerlaw spectrum with a photon index of $\Gamma = 1.8$. 

\subsection{HI properties}

Only low-resolution \HI\ data are currently available for the galaxy ESO\,130-G012 and its environment. These were taken with the Parkes 21-cm multibeam receiver system as part of the \HI\ Parkes All Sky Survey \citep[HIPASS,][]{Barnes2001} and the \HI\ Parkes Zone of Avoidance survey \citep[HIZOA,][]{Donley2005, Donley2006}, which is 5$\times$ deeper than HIPASS. ESO\,130-G012 is catalogued as HIZOA J1222--58 \citep[\vsys\ = $1531 \pm 6$\kms,][]{Stale2016} showing a distinct double-horn \HI\ profile (see Figure~\ref{fig:HI} top). Our measured and derived properties are listed in Table~\ref{tab:properties}. The galaxy's \HI\ mass of \MHI\ = $5.3 \times 10^8$\Msun\ suggests an \HI\ diameter of only \DHI\ $\sim$13~kpc, based on the \MHI-\DHI\ relation \citep{Wang2016}, i.e. only slightly larger than the stellar disk. 

Interestingly, the red-shifted, south-eastern (SE) side of the global \HI\ spectrum is somewhat brighter than the blue-shifted, northwestern (NW) side. Exploring the Parkes \HI\ channel maps we find this asymmetry may be caused by extraplanar \HI\ emission (\MHI\ $\sim$10$^7$\Msun) around 1700\kms, located south-east of the disk, possibly a yet unidentified, interacting dwarf galaxy, likely located within 100~kpc of ESO\,130-G012. This is hinted at in Figure~\ref{fig:HI} (bottom), where we show the mean \HI\ velocity field of ESO\,130-G012 and surroundings. Interferometric \HI\ observations are needed to study the galaxy's gas distribution, kinematics, including its rotation and peculiar motions, and environment. \\

We detect two gas-rich galaxies in the surroundings of ESO\,130-G012: (1) HIZSS\,073 \citep[\vsys\ = 1477\kms,][]{Henning2000}, also known as HIPASS 1221--59 and HIZOA J1221--59, and (2) HIZOA J1211--59 \citep[\vsys\ = 1533\kms,][]{Stale2016} at projected distances of $\sim$70\arcmin\ and $\sim$90\arcmin, ie. 344~kpc and 442~kpc, respectively, for a group distance of $D$ = 16.9~Mpc. HIZSS~073's \HI\ mass is $2.7 \times 10^9$\Msun\ \citep{Ryan-Weber2002, Koribalski2004}, $\sim$4$\times$ larger than that of ESO\,130-G012.  DECaPS images reveal the optical counterpart at $\alpha,\delta$(J2000) = $12^{\rm h}\,21^{\rm m}\,34.4^{\rm s}$, --59\degr\,42\arcmin\,17.3\arcsec\ (size $\sim$ 30\arcsec, $PA \sim 130$\degr, inclination $\sim$ 60\degr). Interferometric \HI\ observations may reveal further dwarf companion galaxies in the vicinity of ESO\,130-G012 which would play a role in triggering star formation through tidal interactions.

\subsubsection{Binding energy and HI kinetic energy}
The dynamical mass of ESO\,130-G012 within $R_{\rm HI}$ = 6.5~kpc is \Mdyn\ = 5.3 ($\pm$0.5) $\times 10^{10}$\Msun\ (see Table~\ref{tab:properties}). This suggests, based on the popular spherical top-hat model \citep{Peebles1993}, a virial radius of $R_{\rm vir} \sim 80$~kpc for which the mass density contrast is $18 \pi^2$ and the binding energy is $-1.9 \times 10^{57}$~erg. 
From the galaxy's \HI\ mass, \MHI \ = 5.3 ($\pm$0.8) $\times 10^8$\Msun, and its rotational velocity, \vrot\ = 188 $\pm$ 6\kms, we estimate the corresponding \HI\ kinetic energy as $K_{\rm rot} \simeq M_{\rm HI} \times v_{\rm rot}^2 = 1.9 ~(\pm 0.5) \times 10^{56}$~erg.
This value, together with the estimate of the system binding energy suggests that  gas cannot leave the halo.

\section{Discussion}
\label{sec:discussion}

In the Local Universe, powerful outflows observed above and below the disks of spiral galaxies typically originate from their nuclear regions, driven either by starbursts or AGN. While outflows or filaments are also sometimes seen from the mildly star-forming disks of edge-on galaxies, their heights are typically quite low. Our discovery of a large-scale ($\gtrsim$30~kpc) bipolar outflow from the $\sim$10~kpc-diameter disk of ESO\,130-G012, was unexpected. --- Does the galaxy's low-luminosity AGN play a role\,? Or can star formation and stellar-driven winds explain the width, large height, and symmetric, edge-brightened bubble shape of the bipolar outflow from the disk of ESO\,130-G012\,? And what triggered the outflow\,? \\

From the detected X-ray emission, we infer that there might be low-luminosity AGN in the center of ESO\,130-G012. But no radio jets are detected, neither a nuclear bi-conical outflow, and the galaxy lies on the radio-FIR relation where normal SF galaxies are found. Nevertheless, as the accretion rate of black holes varies on timescales of $\sim$10--100~Myr, prior AGN outbursts are likely. 

On the other hand, the wide outflow base and bipolar bubble morphology of ESO\,130-G012 favors a stellar outburst scenario. While rare, similar large-scale outflows have been observed in a few nearby disk galaxies, traced by X-ray, \Ha, dust and/or radio continuum (RC) emission (see Table~\ref{tab:comparison}, for example in NGC~1371 \citep{Veronese2025}, which also has a low SFR, NGC~1808 \citep{Kane2024}, NGC~3079 \citep{Hodges2020}, NGC~4217 \citep{Heesen2024}, and NGC~5775 \citep{Heald2022}. Several formation scenarios -- star formation, CR winds, AGN bubbles, ... -- have been put forward, able to explain the observed outflow properties. We find that the energies required to launch the radio emission to heights of $\sim$50~kpc can be supplied even by steady, low star formation in the disk over an extended period of time. See our calculation in the Appendix. So, why don't we see such large-scale outflows more often\,? Will deeper, more sensitive observations, e.g., in the radio continuum, reveal many more large-scale outflows\,? Do the galaxies with already known X-shaped, polarized wings \citep[e.g.,][]{Stein2025} harbour large-scale outflows\,? After some 10 million years, the radio emission fades away, so older outflows are not detectable. Obtaining outflow ages (e.g., via spectral index maps) and SF histories will be important to progress our understanding of outflow formation and timescales. \\

\begin{figure*}[htb] 
\centering
    \includegraphics[width=14cm]{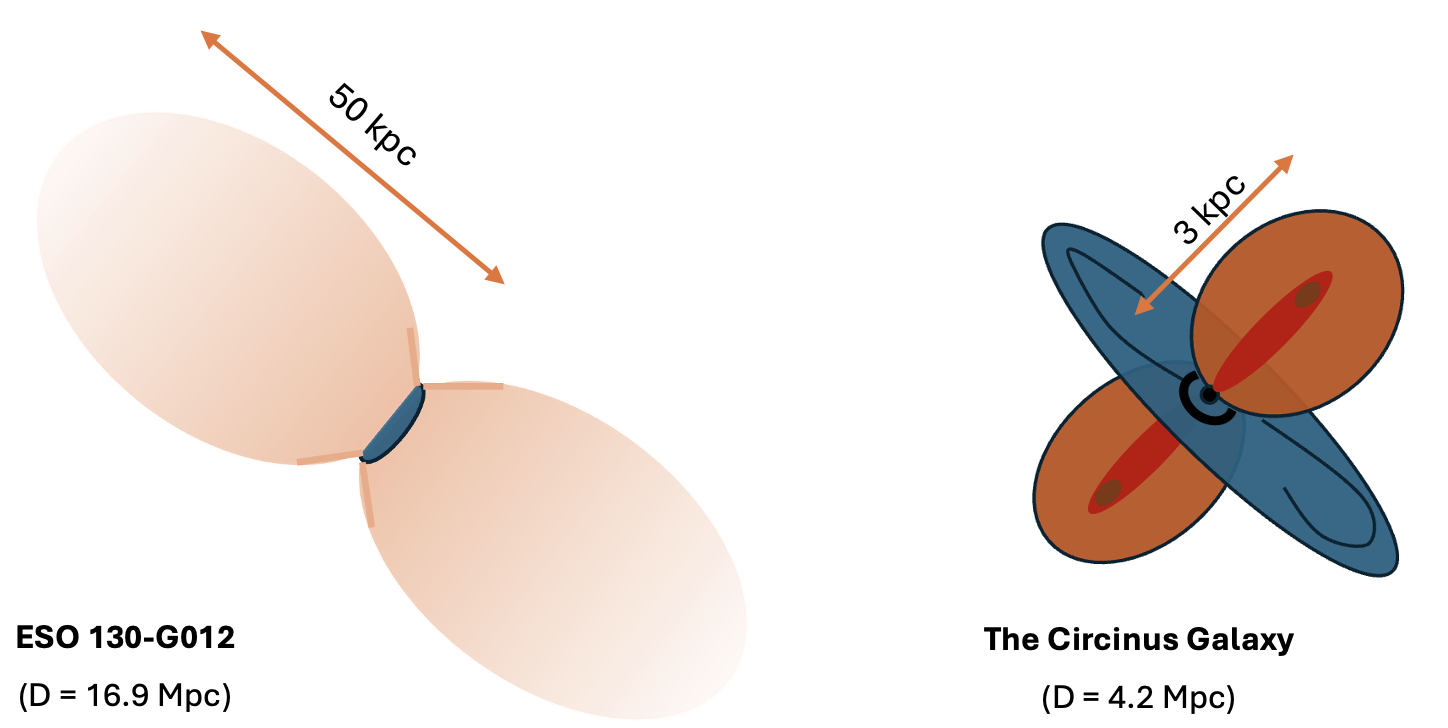}
\caption{Sketches of the bipolar outflows observed in ESO\,130-G012 (left) and the Circinus Galaxy (right, based on \citet{Elmouttie1998-radio}). The galaxy stellar disks, shown in blue color, are indicative.}
\label{fig:sketches}
\end{figure*}

\begin{table}[htb] 
\centering
\begin{tabular}{ccccc}
\hline
 & & Outflow \\
Galaxy Name & Distance & height & comments & References \\
            & [Mpc] & [kpc] \\
\hline
Milky Way & & 8--10 & Fermi bubbles & S10 \\
NGC~253   &  3.2 & 6--12 & \Ha, X-ray, RC & B08, H09 \\
M\,82     &  3.6 & 11 & \Ha, X-ray & DB99, L99 \\
Circinus  &  4.2 & 3 & RC & E98 \\ 
NGC~891   &  10 & $\sim$4.5 & \Ha\ & D90 \\
NGC~1371  & 22.7 & $\sim$20 & RC & V25 \\
NGC~1482  & 22 & 5 & \Ha, X-ray & S04 \\
NGC~1532  & 15 & 10 & RC & M25 \\
NGC~1569  & 3.4  & $\sim$4 & \Ha, X-ray, RC & K10 \\
NGC~1808  & 12.3 & 3 & \HI & K93a,b \\ 
          &  "   & 13 & dust & K24 \\ 
NGC~2992  & 31 & 13.5 & RC, X-ray & V01 \\
NGC~3079  & 19.0 & 30 & X-ray & H20 \\
          &      & 60 & FUV   & H20 \\
NGC~4217  & 20.6 &  7 & \Ha, X-ray, RC & S20 \\ 
           &  "  & 20 & RC & H24 \\
NGC~4383  & 16.5 &  6 & \Ha & W24 \\
NGC~4666  & 27.5 &  9 & \Ha, X-ray & S19 \\ 
NGC~5084  & 29.9 &  9 & X-ray & B24 \\
NGC~5775  & 28.9 & 13 & RC    & H22 \\
NGC~6240  & 107  & 45 & \Ha, X-ray & Y16 \\
Arp~220   & 76   & 15 & X-ray & H96 \\
... \\
Makani    & $z = 0.45$ & 50 & O\,VI & R19, R23 \\
\hline
\end{tabular}
\caption{Nearby galaxies with known large ($\gtrsim$3~kpc) bipolar outflows.}
\label{tab:comparison}
\flushleft References:
B08 \citep{Bauer2008},
B24 \citep{Borlaff2024},
D90 \citep{Dettmar1990},
DB99 \citep{DB1999}, 
E98 \citep{Elmouttie1998-radio}, 
H96 \citep{Heckman1996},
H09 \citep{Heesen2009}, 
H20 \citep{Hodges2020}, 
H22 \citep{Heald2022},  
H24 \citep{Heesen2024},  
K93a \citep{Koribalski1993a}, 
K93b \citep{Koribalski1993b}, 
K10 \citep{Kepley2010},
K24 \citep{Kane2024}, 
L99 \citep{Lehnert1999}, 
M25 \citep{Matthews2025}, 
R19 \citep{Rupke2019}, 
R23 \citep{Rupke2023}, 
S04 \citet{Strickland2004a}, 
S10 \citet{Su2010},
S19 \citep{Stein2019}, 
V01 \citep{Veilleux2001}, 
V25 \citep{Veronese2025},
W24 \citep{Watts2024}, and
Y16 \citep{Yoshida2016}.
\end{table}

\subsection{Comparison with other large-scale galaxy outflows}

In Table~\ref{tab:comparison} we list galaxies with relatively large ($\gtrsim$3~kpc) bipolar outflows. This list includes several well-known starburst galaxies where the observed outflow heights reach 10--15~kpc from the disk \citep[e.g.,][]{Matthews2025,TH2024}. Larger outflows are extremely rare. Notably, in most galaxies, the outflows emerge from their nuclear region. Both the ESO\,130-G012 outflow height and its disk-wide outflow base are highly unusual. See Figure~\ref{fig:sketches} (left) for a sketch of the bipolar outflow from ESO\,130-G012. 

\begin{itemize}
\item Among the lowest observed outflow heights in our sample is the nearby Circinus Galaxy which shows prominent radio bubbles emerging from its active central region \citep[see][and references therein]{Elmouttie1998-radio} extending 3~kpc above and below its disk. The sketch shown in Figure~\ref{fig:sketches} (right) is based on the radio and infrared images of the Circinus Galaxy; we highlight the bipolar radio bubbles and their inner structure, the inner stellar disk, the prominent AGN and the central star-forming ring.  \\

\item The activity and outflow geometry of the starburst galaxy M\,82 is very different from that of ESO\,130-G012. Both \Ha\ and X-ray emission is observed to heights of 11~kpc \citep[e.g.,][]{Adebahr2013}. While exploring the starburst wind from M\,82, \citet{Strickland2000} and \citet{Strickland2009} developed several models, incl. a thin-disk model (their Fig.~6) which produces conical winds with large opening angles, somewhat similar to ESO\,130-G012. \\

\item The nearby, edge-on galaxy NGC~891 -- often considered a twin of the Milky Way -- is well-known for its disk-wide chimneys \citep{Dettmar1990}. Its SFR and \vrot\ are somewhat higher than in ESO\,130-G012, while its \HI\ mass is 10$\times$ than that of ESO\,130-G012. In NGC~891, where SF is detected from the whole disk, \citet{Schmidt2019} find that the best-fitting advection flow is an accelerated galactic wind with midplane velocities of $\sim$150\kms, reaching the escape velocity at a height of 9 -- 17~kpc, depending on the radius. \\

\item In NGC~1532, which is tidally interacting with its smaller elliptical companion, \citet{Matthews2025} find magnetized, highly ordered radio continuum loops originating from star forming regions and extending 10~kpc above and below the disk. They suggest cosmic ray (CR) pressure plays a significant role in launching these outflows. Notably, the north-western radio loops appear to surround the infalling companion, NGC~1531, which is likely responsible for NGC~1532's enhanced SFR of $\sim$2.7\Msun\,yr$^{-1}$. \\

\item The dust filaments and gas outflow from the central starburst region of NGC~1808, emerging perpendicular to its disk, have been known for a long time \citep[e.g.,][]{Koribalski1993a, Koribalski1993b}. More recently, \citet{Kane2024} detected dust filaments extending up to $\sim$13~kpc from the disk in a bi-conical structure. Using the isothermal sheet model, they estimate the potential energy needed to lift the dust and gas to a mean height of $\sim$5~kpc. They find that $5 - 10 \times 10^{56}$~erg is required, which they suggest can be delivered by the SFR of 3.5 -- 5.4\Msun\,yr$^{-1}$ sustained over 4 -- 26~Myr. They also note that the weak AGN in NGC~1808 is energetically not important in driving the observed large-scale outflow. \\

\item The largest outflow that we could find in the literature is detected in the halo of the edge-on, starburst galaxy NGC~3079. \citet{Hodges2020} find an X-shaped outflow reaching 30 and 60~kpc above the disk in X-ray and UV emission, respectively. This is much larger than the inner radio lobes previously seen to emerge from the galaxy's nuclear region \citep{Duric1988,Irwin2003}, which are similar to those in the Circinus Galaxy. The edges of the X-ray outflow in NGC~3079 are found just inside the bubble created by the FUV filaments. \citet{Hodges2020} refer to 3D hydro-dynamical simulations by \citet{Tanner2016} who model the outflow formation within a 1~kpc cube. In \citet{Tanner2025} these simulations are expanded to a 5~kpc box, comparing nuclear starburst and AGN with focus on the X-ray emission. \\

\item In the nearby, edge-on galaxy NGC~4217 \citet{Heesen2024} find a one-sided 20~kpc radio bubble. Using matched LOFAR 144~MHz and JVLA 3~GHz data, they find a spectral gradient (ageing) in the bubble. Fitting CRE advection models to the intensity profiles, \citet{Heesen2024} find the bubble may have been inflated by supernovae over a timescale of 35~Myr. No active galactic nucleus (AGN) is detected in NGC~4217, which has an SFR of 4.6\Msun\,yr$^{-1}$. \\

\item In NGC~4666, \citet{Stein2019} find a box-like 9~kpc radio halo, X-shaped magnetic field structure, and widespread star formation across almost the entire disk. Similar X-shaped polarization structures are also observed in NGC~253 by \citet{Heesen2009} and NGC~5775 by \citet{Tuellmann2000} who explore the formation of such spurs by dynamo action. The SFR of NGC~4666 is 7.3\Msun\,yr$^{-1}$, the highest among the 35 edge-on galaxies in the CHANG-ES sample \citep{Wiegert2015}. \\

\item Interestingly, the mildly inclined spiral galaxy NGC~1371 \citep{Grundy2023, For2021}, which -- like ESO\,130-G012 -- has relatively low SFR and a weak AGN, reveals bipolar bubbles of $\sim$20~kpc height \citep{Veronese2025}. The large \HI\ disk of NGC~1371 resembles that of the Circinus Galaxy \citep{Jones1999, Koribalski2018}, which also hosts an AGN and SF ring, but its radio bubbles are much larger (maybe at a later evolutionary stage) than those in Circinus. \\

\item The "flux tube" model by \citet{Heald2022} for the galaxy-wide outflow in NGC~5775 may be applicable to ESO\,130-G012. They use an outflow base of 14~kpc (this is half the SF disk in NGC~5775), where CRs are advected vertically in the flow of magnetized plasma, expanding adiabatically (their Figure~8). 

\end{itemize}

\subsubsection{Fermi bubbles}
The teardrop-shaped `Fermi bubbles' \citep{Su2010}, which extend 8--10~kpc above and below the Galactic Plane, emerge from the star-forming Galactic Centre region where Sgr\,A$^\star$, our Galaxy's $\sim$10$^6$\Msun\ black hole, resides. The central waist of the elongated bubbles appears to be very narrow ($\sim$100~pc). The bubbles are likely underdense, i.e. filled with hot gas of lower density than the surrounding ISM, expand at $\lesssim$1000\kms\ with energies of $\sim$10$^{55-56}$~erg \citep[e.g.,][]{Crocker2015} and are likely $\sim$10$^7 - 10^8$~yrs old. Magnetic fields may be present, stabilizing the bubble shape and preventing their break-up. Using the Parkes 64-m telescope "Murriyang", \citet{Carretti2013} find two giant, linearly polarized radio lobes, closely corresponding to the Fermi bubbles, with a strong magnetic fields of $\lesssim$15~$\mu$Gauss and suggest they formed via an SF-driven outflow from the Galaxy's central 200~pc \citep[see also][]{Crocker2015}. In their review, \citet[][their Fig.~13]{Sarkar2024} compare simulation snapshots for four different formation scenarios, consisting of an SF-driven wind with SFR = 0.5\Msun\,yr$^{-1}$, an AGN burst, an AGN jet and a tilted AGN jet. 

\subsubsection{Statistics}
The radio outflow from the disk of the nearby edge-on galaxy ESO\,130-G012 was discovered during a visual search of the currently available EMU fields. At the time of writing, the southern sky coverage of EMU was $\sim$50\% complete. This visual search, conducted mainly by BSK for LSB structures such as odd radio circles \citep[ORCs,][]{Norris2021, Koribalski2021}, nearby galaxies \citep{EMU-PS} and radio shell systems \citep[e.g.,][]{Koribalski2024-Physalis} as well cluster relics and halos so far resulted in only one large radio outflow. Such extraplanar radio emission would be recognizable for well-resolved galaxies oriented close to edge-on ($i > 70$\degr), but the detection of bipolar radio bubbles in the nearly face-on spiral galaxy NGC~1371 shows that an edge-on orientation is not necessary. They  are also expected to be gas-rich, likely cataloged in HIPASS \citep{Koribalski2004}, as the gas is fueling the star formation. 

One nearby galaxy with a distinct radio halo detected in the EMU Pilot Survey \citep{EMU-PS} is NGC~7090 (HIPASS J2136--54). It is gas-rich \citep{Koribalski2004} and was previously studied by \citet{Heesen2016}. Another edge-on galaxy with extraplanar emission, detected in EMU, is NGC~5775 (HIPASS J1453+03), which was studied in detail by \citet{Heald2022}. EMU will allow the study of several 100 nearby, edge-on spiral galaxies. For now, we conclude that the huge, bipolar radio bubbles detected in ESO\,130-G012 are very rare, meaning less than one in a 100 edge-on spiral galaxies show such extended extraplanar structures. 

In the sample of $\sim$300 nearby, infrared-bright galaxies studied by \citet{Condon2021-IRAS-BGS} with MeerKAT, only one galaxy (NGC~1532, see above) is noted as having a radio detected outflow.

\subsubsection{Comparison to PN formation}

The hourglass-shaped outflow from the disk of ESO\,130-G012 resembles that of some bipolar planetary nebulae (PNe). Examples include the "Southern Crab" (Hen~2-104) and the "Butterfly Nebula" (NGC~6302). Fast stellar wind from the primary star in a close binary system, whose expansion is funneled by a dense torus of gas, collide with slower winds and produce hourglass-shaped shells, aided by magnetic fields \citep{Blackman2001}. Such magneto-hydrodynamical (MHD) winds from the PNe disks or center can power and shape their bipolar outflows. Magneto-centrifugally driven winds are likely to also play a r\^ole in galaxy outflows. \citet{Balick2019, Balick2023} pose that the injected flows that create the hollow lobes of PNe must be light, “tapered” and injected considerably faster than the lobe expansion speed. They also emphasize the role of toroidal magnetic fields. Light flows produce hollow lobes bordered by dense walls of displaced gas. \citet{Icke2022} propose that the disk at the base of bipolar PNe is itself the source of the outflow.

\subsection{Outflow energetics and timescales}

The enthalpy content of the lobes is $\sim$1 -- $2 \times 10^{56}$ erg for outflow heights of 30 -- 50~kpc (see Appendix). Star formation (SF) at the current rate of 0.1 -- 0.6\Msun\,yr$^{-1}$ can supply 10$^{56}$~erg in $3 \times 10^7$ yr, which is less than the dynamical time estimated as 30~kpc/$c_{\rm s} \sim 3 \times 10^8$ year. For comparison, 10$^{56}$~erg is very similar to the Fermi Bubbles. Even if the structures expand somewhat supersonically and/or less than 100\% of the disk supernova mechanical energy goes into inflating the bubbles, looks like disk SF at the current rate can do it reasonably easily (in $\sim$30 Myr). Interestingly, CRs by themselves are, for fiducial numbers, inadequate (by $\sim$1/2) to power the inflation of the lobes at the sound speed (see our calculations in the Appendix). \\

\begin{figure}
\centering
    \includegraphics[width=8cm]{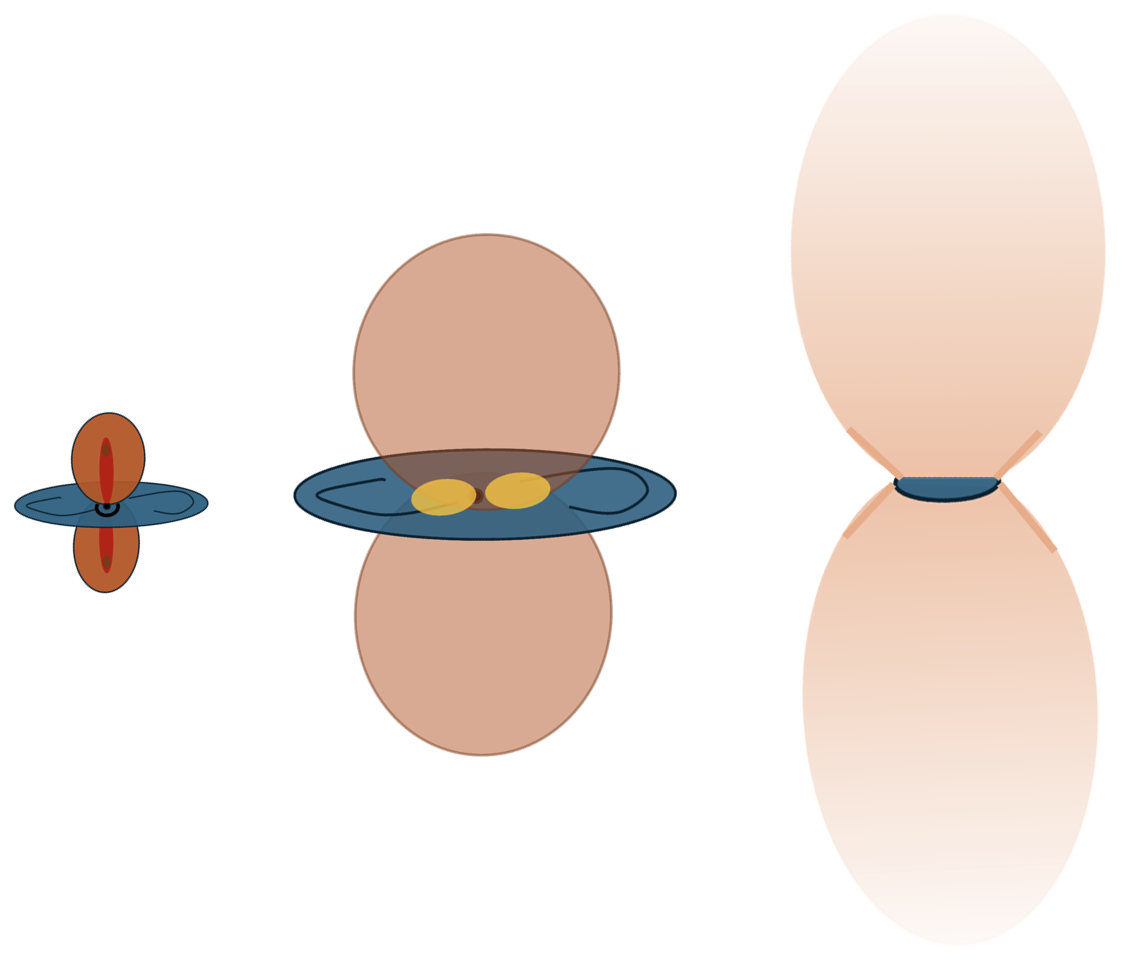}
\caption{Sketches of the bipolar radio continuum outflows from three nearby galaxies, highlighting a possible evolutionary path. From left to right (young to old): Circinus, NGC~1371 and ESO\,130-G012. The sketches are approximate and not to scale.}
\label{fig:sketch-rot}
\end{figure}

Could AGN jets have inflated the observed bipolar lobes of ESO\,130-G012\,? While large double-lobed radio sources are typically hosted by elliptical galaxies, some do have spiral hosts \citep{Mulcahy2016,Koribalski2025-GRGs}. A nearby example is the Circinus Galaxy, whose bipolar radio lobes originate from its central AGN (see Figures~\ref{fig:sketches} \& \ref{fig:sketch-rot}).

A $2 \times 10^7$\Msun\ black hole accreting at the typical rate of 0.1\% of Eddington and with a jet production efficiency of 10\% \citep{TurnerShabala2015} will produce a $3 \times 10^{41}$~ergs\,s$^{-1}$ jet. Because AGN jets couple approximately half their energy to the ambient gas \citep{HardcastleKrause2013, BourneSijacki2021}, over a lifetime of several Myr this provides sufficient energy to inflate the bubbles even in an environment an order of magnitude denser than that adopted in the Appendix. Further circumstantial evidence in favour of an AGN-driven mechanism is the observed ring of star formation in the galactic disk. Simulations by \citet{Dugan2017} showed that strong backflow from jet termination shocks can compress gas in the galactic disk and trigger rings of star formation. AGN winds can inflate similar bubbles and also trigger star formation, however the jets are more efficient (by approximately an order of magnitude because of their mechanical advantage due to being energy-driven \citep{Dugan2014}, see also \citep{CattaneoBest2009}. If the observed radio bubbles arise due to jet activity, faint parsec-scale structures consistent with low lobe luminosity would be expected in VLBI observations -- assuming that no significant modulation in jet power has taken place since the bubble inflation phase. \\

On the other hand, there are no indications of outflow from the galaxy core in ESO\,130-G012, and the observed X-shaped wings forming the base of the bubbles are not seen in AGN-driven outflows. The observed limb brightening suggests that the outflow cones are hollow, possibly filled with hot X-ray gas. The high rotation of the disk may also play a role in terms of generating a dynamo and amplifying the magnetic field into the halo \citep{Stein2025}, while the relatively low mass of ESO\,130-G012 allows outflows to grow larger than in more massive galaxies.

\section{Summary and Conclusions}
\label{sec:conclusions}

The edge-on spiral ESO\,130-G012 is neither a starburst galaxy nor does it host a dominant AGN. Its disk SFR is mild, and a tentative X-ray detection suggests there may be a low-luminosity AGN. Overall, ESO\,130-G012 seems like an unremarkable nearby galaxy located within a loose group, apart from the fact that it hosts one of the largest bipolar outflows known in the Local Universe. Our discovery of a radio continuum outflow reaching 30 -- 50~kpc into the halo on each side of ESO\,130-G012's star-forming disk, which forms the base (waist) of the flow, raises a number of questions. \\  

The shape of the detected bipolar bubbles hold clues as to their origin. Bright X-shaped radio spurs emerge from the boxy thick disk, similar to the polarised radio spurs seen in other galaxies with large outflows. Beyond these bright spurs or "butterfly" wings, the radio emission decreases rapidly, with faint elongated bubbles just detectable on both sides of ESO\,130-G012's stellar disk. In the following, we summarize the galaxy's key properties:
\begin{itemize}
 \item the dynamical mass of ESO\,130-G012, which we derive from its \HI\ rotational velocity (\vrot\ = $188 \pm 6$\kms), is \Mdyn\ = 5.3 ($\pm$0.5) $\times\ 10^{10}$\Msun\ within $R_{\rm HI}$ = 6.5~kpc. The galaxy's \HI\ mass is 5.3 ($\pm$0.8) $\times 10^8$\Msun, i.e. $\sim$1\% of \Mdyn, and its stellar mass is $\sim$1.1 $\times 10^{10}$\Msun, based on the extinction-corrected WISE W1 magnitude.
 \item ESO\,130-G012 is located in a loose group; the galaxy's slightly asymmetric, global \HI\ spectrum suggests it may have a dwarf companion and/or tidal tail.
 \item ESO\,130-G012 has a low disk SFR of $\sim$0.1 -- 0.6\Msun\,yr$^{-1}$. It has a radio core and probably hosts a weak AGN, based on the tentative X-ray detection, but no radio jets are seen. 
 \item ASKAP EMU 944~MHz radio continuum images of ESO\,130-G012 reveal a large-scale hourglass-shaped outflow to a height of at least 30~kpc above and below the disk, possibly extending to $\sim$50~kpc.
 \item the outflow initially rises vertically off the 10~kpc-diameter stellar disk, then broadens laterally while expanding upward (like a wine glass),
 \item bright X-shaped radio wings, likely accompanied by magnetic fields, suggest that the wind is light and hollow, 
 \item the outflow opening angle with respect to the vector orthogonal to the disk is $\sim$30\degr\ on both sides,
 \item ESO\,130-G012's outflow is likely driven by star formation, stellar winds and cosmic rays from the full width of its stellar disk, although we cannot exclude contributions from a more active black hole or central starburst in the past. 
 \item A dynamo induced, poloidal magnetic field might explain the bubble shapes of the observed bipolar outflow.
 \item Our energetics calculations suggest that star formation from the disk of ESO\,130-G012 sustained over 30~Myr can inflate the observed bubbles/lobes. Not yet explained here is the rarity of such large-scale outflows in the Local Universe.
 \item Our discovery of a large-scale radio continuum outflow from the disk of ESO\,130-G012 makes it a promising target to further explore its disk-halo interface and model the outflow formation. This requires a big observing and modeling campaign (see below), which is now under way. Results from this campaign will be presented in follow-up papers.
\end{itemize}

\section{Outlook}

Interferometric \HI\ data as well as wide-band radio continuum and polarization data of ESO\,130-G012 are needed to further examine the observed large-scale radio outflow and its origin. These can be obtained with both the ATCA, which is currently transitioning from CABB \citep{Wilson2011} to an even more versatile correlator (BIGCAT), in L-band (1 -- 3~GHz) and CX-band (4 -- 12~GHz), and MeerKAT extending to lower frequencies. Wide-band radio continuum data are essential to study the magnetic field and spectral index of the galaxy and its bipolar outflow. Particularly, deep ATCA 4-cm (CX-band) data would allow to map the magnetic field in disk of ESO\,130-G012 and its X-shaped winds as the $\lambda^2$-dependent depolarisation is much reduced. ATCA \HI\ maps would provide insights into the galaxy kinematics and any peculiar motions (e.g., halo gas), uncover nearby dwarf companions and tidal interactions.

A spatially resolved spectral index map of the outflow bubbles might not only constrain the time of the outburst event, but also indicate whether the underlying electron population was injected either by SNe or an AGN. Some blazars display very flat spectra with spectral indices lower than the limit achievable through shock acceleration of $\alpha = 0.5$, which would rule out CR injection through SNe. Apart from the spectral properties, multi-frequency radio observations would also constrain the local properties of the CGM by analyzing the rotation measure and fractional polarization across the outflow.

MUSE \Ha\ mosaics of ESO\,130-G012, particularly its thick disk and X-shaped wings would deliver outflow velocities and ionization properties, while deep X-ray observations would likely show the hot gas inside the outflow cones. An analysis of the stellar population of ESO\,130-G012 would provide insights into its star formation history, including possible starbursts in its past. \\


\vspace{-0.3cm}

\section*{Acknowledgments}

We thank Michelle Cluver, Ivy Wong, Ettore Carretti, and Andrew Hopkins for their comments and insightful questions. RMC acknowledges support from the Australian Research Council through the \textit{Discovery Projects} scheme, award DP~230101055, shared with Prof.~Mark Krumholz. RJD acknowledges support by the Deutsche Forschungsgemeinschaft through grant CRC 1491 and the CSIRO ATNF visitors program. 
The Australian SKA Pathfinder is part of the Australia Telescope National Facility (ATNF) which is managed by CSIRO. Operation of ASKAP is funded by the Australian Government with support from the National Collaborative Research Infrastructure Strategy. ASKAP uses the resources of the Pawsey Supercomputing Centre. Establishment of ASKAP, the Murchison Radio-astronomy Observatory (MRO) and the Pawsey Supercomputing Centre are initiatives of the Australian Government, with support from the Government of Western Australia and the Science and Industry Endowment Fund. This paper includes archived data obtained through the CSIRO ASKAP Science Data Archive (CASDA). We acknowledge the Wajarri Yamatji as the traditional owners of the Observatory site. \\


\vspace{-0.3cm}

\section*{Data availability} 

The ASKAP data used in this article are available through CASDA. Image processing and analysis was conducted with the {\sc miriad} software\footnote{https://www.atnf.csiro.au/computing/software/miriad/} and the Karma visualisation\footnote{https://www.atnf.csiro.au/computing/software/karma/} packages.

\appendix 

\section*{Appendix: Lobe enthalpy, expansion time, and energetics}
\label{sec:appendix}

The enthalpy of a bubble or lobe slowly inflated to volume $V_{\rm lobe}$ in pressure equilibrium with its surroundings is
\begin{equation}
\label{eq:enthalpy}
H = \frac{\gamma}{\gamma - 1} ~P_{\rm ext} ~V_{\rm lobe},
\end{equation}
where $\gamma$ is the adiabatic index of the plasma inside the lobe and $P_{\rm ext}$ is the external pressure. For a non-relativistic (NR) monoatomic gas, $\gamma = 5/3$, so $H_{\rm NR} = 5/2 ~P_{\rm ext} ~V_{\rm lobe}$, while  a relativistic (rel) plasma (e.g., a cosmic-ray pair plasma) has $\gamma = 4/3$, so $H_{\rm rel} = 1.6 \times H_{\rm NR}$. The total enthalpy content of {\bf both} lobes is then $H_{\rm tot,NR} = 2 \times H_{\rm NR}$ and $H_{\rm tot,rel} = 3.2 \times H_{\rm NR}$. \\

Approximating each lobe of ESO\,130-G012's outflow as a triaxial ellipsoid with axes $a = 50$~kpc and $b = c = 30$~kpc (see Figure~\ref{fig:bubble-overlays}), gives $V_{\rm lobe} = \frac{\pi}{6} \ a \ b \ c \simeq 6.9 \times 10^{68}$~cm$^3$. \\

Next, we estimate the external pressure, $P_{\rm ext}(h)$, at height $h$, by the halo of ESO\,130-G012. For a fully ionised gas, the total particle number density is $n_{\rm tot} \simeq 2~n_e$ and 
\begin{equation}
 P_{\rm ext}(h) \simeq 2~n_e(h) ~k_{\rm B} ~T_{\rm vir}, 
\end{equation}
where $k_{\rm B}$ is the Boltzmann constant and $T_{\rm vir}$ the virial temperature. For an isothermal halo with \vrot\ = 188\kms, we get:
\begin{equation}
T_{\rm vir} \simeq \frac{\mu ~m_{\rm p} ~v_{\rm rot}^2}{2~k_{\rm B}} \simeq 1.3 \times 10^6 \ {\rm K},
\label{eq:tvir}
\end{equation}
with $\mu \simeq 0.59$ the mean molecular weight and $m_{\rm p}$ the proton mass. This implies an adiabatic sound speed in the halo gas of
\begin{equation}
c_{\rm s} = \sqrt{\frac{\gamma ~k_{\rm B} ~T_{\rm vir}}{\mu ~m_{\rm p}}} \simeq 170\ {\rm km\,s^{-1}} \left(\frac{T_{\rm vir}}{1.3\times10^6\ {\rm K}}\right)^{1/2}
\end{equation}
and an expansion time for each lobe of $t_{\rm exp,s} \simeq a / c_{\rm s}$. For lobe heights of $a$ = 15, 30 and 50~kpc, we get $t_{\rm exp,s} \simeq 0.9$, 1.7, and $2.8 \times 10^8$~yr. \\ 


For the enthalpy calculation, we also need an estimate of the halo gas density. We adopt a representative electron density of $n_{e,50} = 1.5 \times 10^{-4}\ {\rm cm^{-3}}$ at a halo height of $h = 50$~kpc,
%
%
and assume a power-law density profile
\begin{equation}
n_e(h) = n_{e,50} \left(\frac{h}{50\ {\rm kpc}}\right)^{-3/2},
\label{eq:nemodel}
\end{equation}
i.e. $n_e \simeq 9.1$ and $3.2 \times 10^{-4}\ {\rm cm^{-3}}$ for lower heights of $h$ = 15 and 30~kpc, respectively. 
%
Given the density model and
taking $T_{\rm vir} \simeq 1.3\times10^6\ {\rm K}$ (see Eq.~3), we then get $P_{\rm ext} \simeq 3.3$, 1.1 and $0.5 \times 10^{-13}\ {\rm erg\,cm^{-3}}$ for $h$ = 15, 30 and 50~kpc. 
For future use (see \autoref{fig:plotStreamLength}), let us declare here a characteristic equipartition magnetic field amplitude at height $h$ in kpc: 
\begin{equation}
\frac{B^2_{\rm eq,h}}{8 \pi} \equiv P_{\rm ext}(\rm h).
\label{eq:Beq}
\end{equation}

In particular, we find $B_{\rm eq,30} = 1.7~\mu$G and $B_{\rm eq,50} = 1.1~\mu$G given our temperature estimate and density model. \\

Substituting $P_{\rm ext} = 0.5 \times 10^{-13}$~erg\,cm$^{-3}$ and $V_{\rm lobe} = 6.9 \times 10^{68}$~cm$^{3}$ we get $H_{\rm tot,NR} \simeq 2.0 \times 10^{56}$~erg to reach an outflow height of 50~kpc \citep[see also][their Section~4.2]{Heesen2024}.
This is a lower limit to the total energetics: it neglects the fact that the bubbles have expanded through denser gas at lower galactic elevations; that there may still be significant kinetic energy in the outflowing material (in the case of supersonic expansion, in particular), radiative losses from shocked matter (again, in the case of supersonic expansion); that work has been done against the gravitational potential in lifting the material of the bubbles; and it does not count the energy in their magnetic field or cosmic ray content. See Table~\ref{tab:theory} for a summary of the adopted and derived values. \\

\begin{table*}[] 
\centering
\begin{tabular}{ccccccccc}
\hline
\multicolumn{3}{c}{galaxy outflow} & expansion time & electron density & external pressure & enthalpy & mechanical power  \\
height & width & $V_{\rm lobe}$ & $t_{\rm exp,s}$ & $n_{\rm e}$ & $P_{\rm ext}$ & $H_{\rm tot,NR}$ & $\dot{E}_{\rm mech,s}$ \\
 kpc] & [kpc] & [10$^{68}$ cm$^3$] & [10$^8$ yr] & [10$^{-4}$ cm$^{-3}$] & [10$^{-13}$ erg\,cm$^{-3}$] & [10$^{56}$ erg] & [10$^{40}$ erg\,s$^{-1}$]\\
\hline
  15   & 9 & 0.2 & 0.9 & 9.1 & 3.3 & 0.3 & 1.2 \\
  30   & 20 & 1.8  & 1.7 & 3.2 & 1.1 & 1.1 & 2.1 \\
  50   & 30 & 6.9  & 2.8 & 1.5 & 0.5 & 2.0 & 2.3 \\
\hline 
\end{tabular}
\caption{Calculations for different galaxy outflow heights assuming $T_{\rm vir} = 1.3 \times 10^6$~K and $c_{\rm s}$ = 170\kms. }
\label{tab:theory}
\end{table*}

We now study the source of energy that drives the expansion of the gaseous lobes (SF) and the amount of stellar mass formed during expansion. The total energy implied by the above estimates is supplied over some inflation timescale $t_{\rm infl}$ by some astrophysical processes that liberate a mechanical power, $\dot{E}_{\rm mech}$, occurring in the disk of ESO\,130-G012: $H_{\rm tot} = \dot{E}_{\rm mech} \ t_{\rm infl}$.
First, as a rough lower bound on the power requirements, we assume that the lobes expand at the sound speed. For outflow heights of 30~kpc and 50~kpc this requires $\dot{E}_{\rm mech,s} \sim H_{\rm tot} / t_{\rm exp,s}$ = $2.1 \times 10^{40}$~erg\,s$^{-1}$ and $2.3 \times 10^{40}$~erg\,s$^{-1}$, respectively. 

\section*{SF energy from ESO\,130-G012}
We compare the above energy requirements to the mechanical power available from star formation, $\dot{E}_{\rm SF}$, at the current rate of SFR $\sim$ 0.1 -- 0.6\Msun\,yr$^{-1}$, with $10^{51}$~erg of kinetic energy injected into the ISM per core collapse supernova and around 100\Msun\ of total stellar mass formed to generate a single supernova \citep[e.g.,][]{Roth2021}:
\begin{eqnarray}
    \dot{E}_{\rm SF} \simeq (0.3 - 1.9) \times 10^{41} \ {\rm erg  \ s^{-1}}~.
\end{eqnarray}

This implies that star formation at the current rate can inflate the lobes provided they do not inflate a lot faster than the halo sound speed. Saying this differently, star formation can inflate the lobes at an average speed with a Mach number up to $\mathcal{M} \sim$ few. A total stellar mass of $\sim$2.8 $\times 10^8 \ {\rm yr} \ \times \ (0.1-0.6) \ \rm M_\odot \ {\rm yr^{-1}} \simeq (0.3-2) \times 10^8$\Msun\ would form over the lobes' inflation. Even at the upper end, this is less than the current \HI\ mass in the galaxy. This scenario requires that star formation be sustained over a timescale considerably smaller than the depletion timescale (see \autoref{tab:properties}). \\

Interestingly, for fiducial numbers -- in particular, adopting that around 10\% of the mechanical energy liberated by star formation is processed into CRs via ISM shocks at accompanying supernova remnants \citep[e.g.,][and references therein]{Roth2021} -- we see that CRs accompanying star formation do not seem quite adequate by themselves to inflate the lobes:
\begin{eqnarray}
    \dot{E}_{\rm CR, SF} \simeq  0.1 \ \dot{E}_{\rm SF} 
    \simeq (0.3 - 1.9) \times 10^{40} \ {\rm erg \ s^{-1}}~. 
\end{eqnarray}

\subsection*{Cosmic ray electron constraints}

\begin{figure}[htb] 
\centering
    \includegraphics[width=7cm]{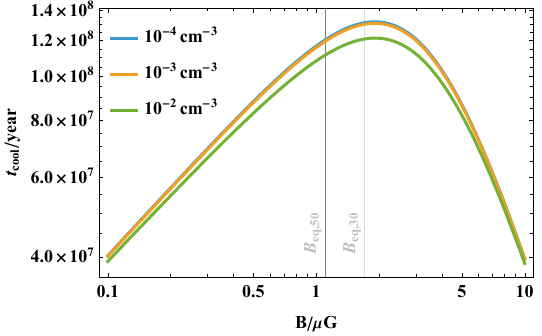}
\caption{The cooling time for electrons with energy $E_{\rm 944}$ as a function of magnetic field amplitude. The vertical grey lines denote the magnetic field amplitudes that would be in approximate equipartition at 30 and 50~kpc, respectively, with $P_{\rm ext}$ given our density and temperature assumptions. Note that the equipartition field amplitudes are close to the magnetic field that maximises the lifetime of CR electrons synchrotron radiating at 944~MHz.}
\label{fig:plotCoolingTime}
\end{figure}

The characteristic energy of cosmic ray electrons synchrotron emitting at 944~MHz in a magnetic field amplitude of $B_{\mu} \equiv B/(\mu G)$ is:
\begin{equation}
    E_{944} \simeq 9.4 \ {\rm GeV} \ B_{\mu}^{-1/2}
\end{equation}
The cooling time of electrons with this characteristic energy is shown in \autoref{fig:plotCoolingTime} for three different number densities $n_{\rm H}$. Here we account for ionisation, bremsstrahlung, synchrotron, and inverse Compton (IC) losses; IC losses are assumed to occur on the cosmic microwave background (CMB).
Bremsstrahlung and ionisation losses are highly subdominant at $E_{944}$ for number densities typical of the halo gas.
Note that there is a definite maximum lifetime for $B \sim 2 \ \mu$G which 
is less than half the expansion timescale of each lobe at the sound speed.
CR electrons, injected in the plane of the galaxy, can move a certain distance given this timescale and depending on the transport mechanism. 
In the limit that CR electrons are `frozen-in' and do not move appreciably with respect to the thermal plasma, the fact that we detect synchrotron emission
up to the full $\sim$50~kpc extent of the lobes would then imply that either the lobes are expanding at $\gtrsim$ 2$ c_{\rm s}$ (so there should be shocks at the edges of the lobes, albeit not necessarily strong ones) or there are mechanisms injecting \citep[e.g., via hadronic collisions $pp \to X \ e^\pm$;][]{Crocker2015} or re-accelerating CRs close to the lobe edges \citep[e.g.,][]{Mertsch2019}.

However, this `frozen-in' limiting behaviour is unlikely to be realised in practice: even in the event that the lobe edges are expanding at the sound speed or slower, there may be faster internal flows of the plasma filling the lobes capable of advecting the cosmic rays and, even in the absence of such coherent internal flows, the CRs can be expected to move with respect to the plasma due to some combination of streaming and/or diffusion. \\

\begin{figure}[htb] 
\centering
    \includegraphics[width=7cm]{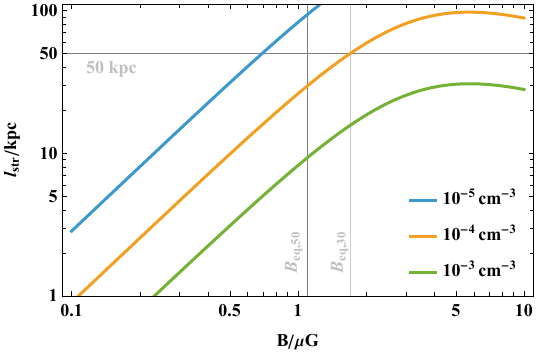}
\caption{The maximum streaming distance for electrons with energy $E_{\rm 944}$ with a streaming speed equal to the Alfv\'en speed for the given magnetic field and the nominated proton number density cooling time for electrons with $E_{\rm 944}$ as a function of magnetic field amplitude. The horizontal line corresponds to the full 50~kpc height of each lobe and the vertical grey lines denote the magnetic field amplitudes that would be in approximate equipartition with $P_{\rm ext}$ given our density and temperature assumptions.}
\label{fig:plotStreamLength}
\end{figure}

(1) If CR electrons of $E_{944}$ stream 
at speed $v_s$ along combed magnetic field lines at a speed equal to the Alfv\'en speed, $v_{\rm s} \to v_{\rm A} (B_\mu, n_H)$, they can reach a distance, $l_{\rm str}$, shown by the curves in Figure~\ref{fig:plotStreamLength} with
\begin{equation}
    l_{\rm str}  
    = v_A(B,n_H) ~t_{\rm cool}(E_{944});
\end{equation}
were streaming at the Alfv\'en speed the only important transport process and, in addition, were there no process of {\it in situ} CR injection or re-acceleration, then we would have the following constraints on the conditions (cf.~\autoref{fig:plotCoolingTime}): $B \gtrsim 0.5 \ \mu$G and $n_H \lesssim 10^{-4}$ cm$^{-3}$.  
These putative constraints are not, in fact, unreasonable; the lower limit on the magnetic field is close to $B_{\rm eq,50}$ and the density constraint is also not unreasonable if the lobes are filled by supernova heated gas: given a total mass of stars formed over the inflation of the lobes $(0.3-2) \times 10^8$\Msun,  were a similar total mass of hot gas lofted into the lobes \citep[e.g.,][]{Crocker2012}, this would correspond to $n_H$ in the range $(0.3-2)\times 10^8 \ M_\odot/(2 \times 6.9 \times 10^{68} \ {\rm cm}^3 \ m_p) \sim (0.3-2) \times 10^{-4}$ cm$^{-3}$.

However, these constraints are necessarily relaxed under many plausible circumstances:
i) in the halo gas, Alfv\'en wave damping processes may mean that $v_s \gg v_A$ \citep[e.g.,][]{Wiener2013};
ii) there may be a non-negligible contribution of advective motion due to plasma bulk motion that adds to the electrons' effective velocity with respect to ESO 130-G012 (such bulk motion could be either coherent internal plasma flows within the lobes' slowly expanding edges -- necessitating back flows and/or gas pile-up inside the lobe edges -- or simply the overall expansion of the lobes); 
and iii) there may be {\it in situ} re-acceleration processes, as already mentioned. \\

\begin{figure}[htb] 
\centering
    \includegraphics[width=7cm]{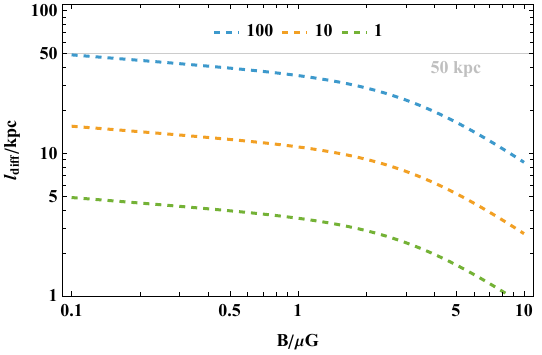}
\caption{The maximum diffusion distance for electrons with $E_{\rm 944}$ with a diffusion coefficient given by \autoref{eq:diffusion} and a cooling time as shown in \autoref{fig:plotCoolingTime}. 
The different colours are for different values of $\chi$ as denoted in the legend.
The horizontal line corresponds to the full 50~kpc height of each lobe and the vertical grey lines denote the magnetic field amplitudes that would be in approximate equipartition with $P_{\rm ext}$ given our density and temperature assumptions.}
\label{fig:plotDiffLength}
\end{figure}

(2) CR electrons of energy $E_e$ which diffuse with some diffusion coefficient parameterized, following \citet{Gabici2007}, as
\begin{equation}
\kappa(E_e,B) =  5.2 \times 10^{27} \ \chi \ {\rm cm^2 s^{-1}} \left(\frac{E_e}{\rm GeV} \right)^{1/2} \left(\frac{B}{ \mu {\rm G}} \right)^{-1/2}
    \label{eq:diffusion}
\end{equation}
(where the case $\chi = 1$ approximately reproduces the diffusion coefficient inferred for CRs in the nearby plane of the Milky Way), would reach a maximum distance given approximately by
\begin{equation}
    l_{\rm diff}(E_e,B) \simeq \left(2 \kappa(E_e,B) ~t_{\rm cool}( E_e) \right)^{1/2};
\end{equation}
we show $l_{\rm diff}(E_e,B)$ for the case $E_e = E_{\rm 944}$ in \autoref{fig:plotDiffLength} where the different colours correspond to $\chi \in \{1,10,100\}$.
Purely diffusive transport would evidently constrain the lobe fields to be weak and the effective diffusion coefficient to be $\gtrsim$ 100 the in-plane Milky Way value at $\sim$GeV energies; such a high diffusion coefficient need not be unreasonable \citep[e.g.,][]{Butsky2023} but, as for the streaming case, there are many plausible ways to relax these constraints.

\printendnotes

\bibliography{E130-Outflow} 

\end{document}